\documentclass{article}
\usepackage[margin=1in]{geometry}
\usepackage{amsmath}
\usepackage{amssymb}
\usepackage{graphicx}
\usepackage{xcolor}
\usepackage{booktabs}
\usepackage{longtable}
\usepackage{natbib}
\usepackage{hyperref}
\hypersetup{colorlinks=true, citecolor=blue, linkcolor=blue, urlcolor=blue}
\usepackage{lineno}      

\title{Magnetospheres, Habitability and the Radio Sky: A Kinetic Radiometric Bode's Law Forecast for Giant Exoplanets}

\author{Bi-Xuan Zhao$^1$, Hui Zhang$^1$\thanks{Corresponding author: zhangh@shao.ac.cn}, Xiao-Long Yang$^1$, Quan Guo$^1$\\
$^1$Shanghai Astronomical Observatory, Chinese Academy of Sciences,\\80 Nandan Road, Shanghai 200030, China}
\date{}

\begin{document}

\maketitle

\begin{abstract}

The habitability of exoplanets and their long-term atmospheric retention can be influenced by the host stellar wind. Electron–cyclotron radio emission generated by magnetosphere--stellar wind coupling represents one of the few direct observational diagnostics of such interactions. Nevertheless, no confirmed radio detection of an exoplanet has been reported to date. In this work, we apply the kinetic Radiometric Bode’s Law to giant exoplanets to predict their characteristic frequencies $\nu_{p}$ and flux densities $S_{\nu}$, with comparisons against the terrestrial ionospheric cutoff and the capabilities of existing and planned radio arrays. The predicted frequency distribution depends strongly on the adopted rotation prescription. Under the fiducial model, which assumes tidal locking for close-in planets, 80.1\% of the sample has characteristic emission frequencies below 10 MHz. This fraction decreases to 52.7\% when all planets are assigned Jupiter's rotation rate. These results motivate complementary ground- and space-based low-frequency searches, while showing that the relative importance of the sub-ionospheric window remains model dependent. Targets with high $\nu_{p}$ are potential priorities for deep LOFAR/SKA-class campaigns, but ranking also depends on predicted flux, distance, beaming, stellar contamination, and facility sensitivity. Characterizing this exoplanet radio landscape is a necessary step toward assessing whether magnetospheric detections can place constraints on planetary habitability.

\end{abstract}


\section{Introduction}

Since the discovery of 51~Pegasi~b \citep{Mayor1995}, more than 6000 exoplanets have been confirmed (Sep, 2026. NASA Exoplanet Archive\footnote{\url{https://exoplanetarchive.ipac.caltech.edu}}). Transit and radial-velocity surveys now deliver masses, radii, and increasingly, atmospheric characterizations \citep{Mayor2003,Ricker2015,Lo2015,Stassun2018,Seager2007,Charbonneau2007}. Intrinsic fields can deflect stellar-wind plasma, suppress non-thermal escape, and shield surface environments from energetic particles. These effects are especially acute around active M dwarfs, whose winds and flares can otherwise strip or irradiate close-in worlds \citep{Griemeier2005a,Lammer2007,Vidotto2013,Driscoll2015}. Habitability assessments that ignore magnetospheric coupling are therefore incomplete. Direct field measurements for exoplanets remain scarce. Low-frequency radio emission from magnetosphere--stellar-wind interactions offers one of the few observational paths into that physics.

Exoplanets with intrinsic magnetic fields \citep{Brain2024,Callingham2024} are expected to generate low-frequency radio emission through interactions between their magnetospheres and the stellar winds of their host stars \citep{Farrell2003,Ip2004}. A direct analogy is observed for Solar System planets \citep{Zarka2001}. Jupiter efficiently accelerates magnetospheric electrons through the electron--cyclotron maser instability (CMI) with its strong magnetic field, producing bright non-thermal radio bursts \citep{Zarka1998a,Zarka2004}. For exoplanets with strong magnetic fields, similar CMI-driven processes are expected to produce detectable radio emission \citep{Zarka2018}. The characteristic frequencies are set by the cyclotron frequency and thus by the surface field strength \citep{Wu1979,Melrose1982,Zarka1998a,Zarka2007,Callingham2024}. The emitted power is commonly estimated from stellar-wind energetics using the kinetic Radiometric Bode's Law (RBL; \cite{Desch1984,Farrell1999,Lazio2004,Zarka2007}) or related radio-magnetic scalings \citep{Zarka2007,Griemeier2005a}. The interaction between the planetary magnetic field and stellar-wind particles can accelerate electrons at the magnetopause or in the magnetotail, generating non-thermal radio emission \citep{Louarn1992,Zarka1998a,Louis2025}. Stellar systems may also host super-Jupiters whose intense magnetic fields, sustained by vigorous convection and rapid rotation, can rival or even exceed the host-star radio emission under favorable conditions \citep{Showman2010,Carson2013,Dempsey2021,Ashtari2022}. Detecting such signals at stellar distances remains extremely challenging \citep{Griemeier2005a}. Jupiter's anisotropic beaming can encode rotation and spin--orbit locking \citep{Zarka2004,Zarka2019}. If placed at typical exoplanet distances, Jupiter-like emission would still be relevant for sensitive low-frequency arrays. This contrast with optical/infrared planet-to-star contrasts is a main reason radio searches have focused on magnetized giants.

The Radiometric Bode's Law (RBL) is an empirical scaling calibrated on Solar System magnetized giants \citep{Desch1984,Farrell1999,Lazio2004,Zarka2007}. Extending this relation across the full exoplanet census is therefore an extrapolation whose uncertainties are greatest for low-mass planets, where Jupiter-analogue magnetospheres, and thus magnetic protection of atmospheres, are least secure, and where the $(R/R_{\mathrm{J}})^{-3}$ factor in the kinetic flux formula can dominate the predicted $S_{\nu}$ \citep{Farrell1999,Lazio2004}. Close-in giant planets remain the most natural targets for kinetic RBL forecasts. They occupy the regime in which a giant-planet RBL extrapolation is considered more plausible than for terrestrial or ice-giant analogues, and they experience strong stellar-wind forcing, which is relevant to both radio power and atmospheric erosion. They also have driven much of the theoretical and observational literature \citep{Zarka2007,Griemeier2007,Lazio2004}. \citet{MousaviSadr2025} applied the kinetic RBL to 1330 confirmed exoplanets and selected SKA-accessible targets with machine-learning diagnostics. Relative to catalog-update approaches, our contribution is physical rather than purely demographic. We restrict the primary analysis to a mass-selected giant-planet sample ($M_{p}>0.1\,M_{J}$) without treating this cut as a Solar System-validated applicability boundary, exclude NASA Exoplanet Archive upper-limit masses rather than treating flagged limits as measured masses, quantify how tidal locking versus a fixed Jovian spin prescription shifts the $\sim$10 MHz pile-up, and compare facility frequency windows, including sub-ionospheric access, as diagnostics of which magnetospheres can be probed.

Despite decades of searches with UTR-2, VLA, GMRT, LOFAR, MWA and related facilities, no exoplanetary radio detection has been confirmed \citep{Zarka2001,Lazio2004,Sirothia2014,Turner2021}. Candidate signals (e.g., toward $\tau$~Bootis, Proxima Centauri and YZ~Ceti) remain contested or unconfirmed, often limited by variability or stellar contamination \citep{Turner2021,Elekes2023,Pineda2023}. Two structural obstacles dominate. The terrestrial ionosphere blocks frequencies $\lesssim$ 10 MHz \citep{Zarka1997,Griemeier2007,Turner2019}. Galactic background noise degrades low-frequency sensitivity \citep{Zarka2007}. Published flux limits spanning microjanskys to janskys are compiled in Table~\ref{tab:searches}. They illustrate both improving facility reach and the continuing non-detections. These non-detections motivate clearer population-level forecasts of where RBL emission should appear in $(\nu_{p},S_{\nu})$ space, and which modeling assumptions most affect that forecast. Near-term strategies include deep, polarization-capable monitoring of the brightest predicted targets with LOFAR/SKA-class arrays at accessible frequencies. Longer term, space- or lunar-based arrays are needed to reach below the ionospheric cutoff \citep{Lazio2004,Burns2019,Chen2024}.

Here we address a population-level question that links magnetospheric physics to survey design and, ultimately, to habitability diagnostics. For a mass-selected giant-planet sample ($M_p > 0.1\,M_J$; $N = 898$), we investigate the predicted distributions of characteristic emission frequencies and flux densities and assess their implications for probing planetary magnetic environments and atmospheric protection. Using a consistently defined mass-selected sample for population statistics and a separately filtered subsample for target prioritization, we map the $(\nu_p, S_\nu)$ distribution, examine the effects of tidal locking relative to fixed Jovian rotation, test the robustness of target rankings using a subset with catalog masses that are not upper limits and mass perturbations, and assess nominal frequency coverage and model-flux exceedance for representative ground- and space-based facilities, including access to frequencies below the terrestrial ionospheric cutoff.

This paper is organized as follows: Section~\ref{sec:methods} defines the giant-planet mass-selected sample and the kinetic RBL framework, including the physical assumptions that control $\nu_{p}$ and $S_{\nu}$. Section~\ref{sec:results} presents the population frequency distribution, its dependence on rotation, the facility-level implications for magnetospheric detection, and the contrast with planets below that mass cut. Section~\ref{sec:discussion} discusses limitations and the relation to prior population RBL studies. Section~\ref{sec:conclusions} summarizes the scientific conclusions for magnetospheric detectability and habitability-motivated survey strategy.

\section{Sample and Methods}
\label{sec:methods}

\subsection{Sample selection}

Our goal is not an all-sky catalog exercise but a mass-selected forecast. We apply the kinetic RBL to the regime in which a giant-planet extrapolation is considered more plausible than for lower-mass worlds, then ask what that implies for detecting the magnetospheres that can shield atmospheres. We therefore start from confirmed exoplanets in the NASA Exoplanet Archive (NEA; \cite{Christiansen2025}; retrieved in 2026 September), and we adopt the default parameter values. We retain planets with the distance, orbital period, and radius needed to evaluate Equations~\eqref{eq:Snu}--\eqref{eq:nup}. Semi-major axes are taken from the catalog when available and otherwise inferred from Kepler's law. We retain $P_{\mathrm{orb}}<5000$\,d so that $a$ remains adequately constrained, while still spanning hot, warm, and cold giant regimes.

Mass assignment for the  $M_{p}>0.1\,M_{J}$ cut follows two mutually exclusive cases, using the NEA limit flags $\texttt{pl\_bmassjlim}$ and $\texttt{pl\_massjlim}$ , where a flag value of 1 indicates an upper limit. Case (i) adopts a non-limit measured mass if a numeric $\texttt{pl\_bmassj}$ or $\texttt{pl\_massj}$ is available and is not flagged as an upper limit. Case (ii) estimates $M_{p}$ from radius via the median broken power-law mass-radius relation of \cite{Chen2017} if neither column supplies a non-limit measured mass and the object is not upper-limit-only. Upper-limit-only objects have a numeric catalog mass flagged only as an upper limit and are excluded. Their limit is neither treated as a measured mass nor replaced by a \cite{Chen2017} estimate, so they cannot satisfy the mass cut. HIP~67522 c has limit flags but no numeric mass entry and therefore falls under case (ii). Applying these rules to the September 2026 NEA default-parameter download used here yields 
N = 898, of which 112 use case (ii), or $\sim 12.5\%$. Because the cyclotron frequency scales steeply with mass ($\nu_{p}\propto M^{5/3}$ in Equation~\eqref{eq:nup}), mass uncertainties translate directly into uncertainty on the magnetic field strength relevant to habitability assessments. We retain the estimated-mass subset to avoid a selection against transit discoveries, but treat it as a systematic uncertainty rather than as precision magnetospheric targets. As a robustness check we also recompute population statistics and the Top-20 ranking on the catalog non-upper-limit mass subset alone ($N=786$; case~(i) only; an unset NEA limit flag, not an independent literature confirmation; Section~\ref{sec:results}). The catalog non-upper-limit mass subset is defined solely by the NEA mass-limit flags and should not be interpreted as a sample of independently literature-verified mass measurements.

The primary cut is a working mass selection, not a claim that the RBL has been validated for all objects above that threshold. The kinetic RBL is calibrated on magnetized Solar System giants \citep{Desch1984,Farrell1999,Lazio2004}; extending it unchanged to terrestrial or ice-giant analogues assumes a Jupiter-like magnetosphere that may not exist. We therefore restrict the primary analysis to \(M_{p}>0.1\,M_{J}\), above the ice-giant / gas-giant transition commonly adopted in exoplanet demography (\(\sim\)30\,\(M_{\oplus}\)), that is, the regime in which a giant-planet RBL extrapolation is considered more plausible. This cut does not certify applicability for high-mass brown-dwarf candidates, young inflated giants, or systems with atypical hosts, and those objects remain extrapolations within the mass-selected sample. White-dwarf hosts are identified in the parent download by either of two conditions, applied before any missing-radius substitution: a catalog stellar radius \(R_\star<0.05\,R_\odot\), or a planet name beginning with WD. The radius threshold lies in the gap below the next smallest catalog radius (\(0.091\,R_\odot\)). The name condition recovers a host whose stellar radius is absent from the catalog. The September 2026 download returns four rows. Three meet the radius cut: ZTF~J1230-2655\,b (\(R_\star=0.0123\,R_\odot\)), WD~1856+534\,b, and ZTF~J1828+2308\,b. WD~0806-661\,b is selected by name. It has no catalog stellar radius and no orbital period, so it never enters the working sample. ZTF~J1230-2655\,b has both NEA mass-limit flags equal to 1 and is already outside the \(N=898\) sample under the upper-limit-only rule. The two white-dwarf hosts that remain inside \(N=898\) are removed from the priority ranking rather than left at its head. WD~1856+534\,b is stored as $M_{p}=13.8\,M_{J}$ with both NEA limit flags equal to 0, but \citet{Vanderburg2020} report $13.8\,M_{J}$ as a 95\% confidence upper limit from the non-detection of thermal emission, not as a measured mass. Dropping only this object from the \(N=898\) totals leaves \(N=897\). The number of planets with \(\nu_{p}<10\)\,MHz stays 719, and the median \(S_{\nu}\) stays \(6.4\times10^{-5}\)\,Jy. The median \(\nu_{p}\) moves from 1.82 to 1.81\,MHz, the fraction below 10\,MHz from 80.1\% to 80.2\%, and the 16--84th percentile range from 0.30--13.7 to 0.30--13.4\,MHz. Its 691\,MHz prediction lies only in the SKA1-Mid and uGMRT windows, which fall from 13 to 12 and from 10 to 9. Every other in-band count is unchanged. ZTF~J1828+2308\,b orbits a white dwarf; \citet{Parsons2025} measure the companion as a brown dwarf of $19.5\pm0.8\,M_{J}$ (the archive default is $20\,M_{J}$). The solar-calibrated $\dot{M}_{\mathrm{ion}}$--$v_{\infty}$ law is not a ranking basis for either host. Inserting the archive masses into Equations~\eqref{eq:Snu}--\eqref{eq:nup} gives 3.25\,Jy at 691\,MHz for WD~1856+534\,b and 0.859\,Jy at $1.32\times10^{4}$\,MHz for ZTF~J1828+2308\,b. Those values are not used as ranks. The priority ranking applies two cuts on top of the mass-selected sample. The two in-sample white-dwarf hosts are excluded, along with seven hosts with catalog ages below 0.1 Gyr: HIP 67522 b, HIP 67522 c, TOI-837 b, TOI-1227 b, IRAS 04125+2902 b, TOI-6448 b, and NGTS-33 b. The solar-wind age factor $(1+t/\tau)^{-1.86}$ ($\tau=25.6$ Myr) places six of these seven at high $S_\nu$. Missing catalog ages are filled with 4.6 Gyr and are not part of this cut. Five of the seven young hosts have \cite{Chen2017} mass estimates. The nine host/age excluded rows remain in the $N=898$ population totals and are listed in Table 3, but are not assigned priority ranks. In addition, Kepler-415\,b is omitted from the priority ranking \citep{Hadden2014} because its catalog $M$--$R$ pair implies an extreme bulk density driven by a highly uncertain TTV mass; it remains in the $N=898$ totals and is listed with those rows in Table~3. After these cuts, the ranking sample contains $N=888$ planets, while the corresponding catalog non-upper-limit mass subset contains $N=781$. The priority Top-20 (Table 4) contains no \cite{Chen2017} mass estimates.

For contrast only, we also evaluate the same formulae on the full confirmed-exoplanet sample ($N=4428$; Figures 5 and 6). Objects with $M_p\leq0.1\,M_J$ illustrate how the scaling extrapolates below our mass cut and are not treated as equal-footing RBL targets. All population statistics and facility fractions use the giant-planet mass-selected sample ($N=898$), whereas the priority ranking uses the filtered subset ($N=888$).

\subsection{The kinetic Radiometric Bode's Law}

The scientific goal is not merely to establish a flux limit but to constrain magnetic field strength and magnetospheric interaction geometry. Full MHD or particle simulations capture that physics for individual systems, yet they do not scale across heterogeneous catalogs \citep{Zarka2007,Hess2011,Griemeier2017,Vidotto2020}. In the absence of confirmed exoplanetary radio detections, population forecasts therefore still rely on empirical energetics. We adopt the kinetic Radiometric Bode's Law, which connects incident stellar-wind power to radiated radio power. For kinetic-energy-dominated injection the conversion efficiency is $\beta\sim10^{-5}$, whereas magnetically dominated injection yields $\beta\sim10^{-3}$ \citep{Zarka2001}. Under these scalings, hot Jupiters with Jupiter-like fields can radiate $10^{3}$--$10^{6}$ times more radio power than Jupiter \citep{Zarka2014}. The predicted flux density follows \citep{Farrell1999,Lazio2004,Li2023,Narang2024}.

\begin{equation}
\begin{aligned}
S_{\nu} &= 4.6\,\mathrm{mJy}
\left(\frac{\omega}{\omega_{\mathrm{J}}}\right)^{-0.2}
\left(\frac{M}{M_{\mathrm{J}}}\right)^{-0.33}
\left(\frac{R}{R_{\mathrm{J}}}\right)^{-3}
\left(\frac{\Omega}{1.6\,\mathrm{sr}}\right)^{-1}
\left(\frac{d}{10\,\mathrm{pc}}\right)^{-2} \\
&\quad\times
\left(\frac{a}{1\,\mathrm{au}}\right)^{-1.6}
\left(\frac{\dot{M}_{\mathrm{ion}}}{10^{-11}\,M_{\odot}\,\mathrm{yr}^{-1}}\right)^{0.8}
\left(\frac{v_{\infty}}{100\,\mathrm{km}\,\mathrm{s}^{-1}}\right)^{2},
\end{aligned}
\label{eq:Snu}
\end{equation}

\begin{equation}
\nu_{p}=23.5\,\mathrm{MHz}
\left(\frac{\omega}{\omega_{\mathrm{J}}}\right)
\left(\frac{M}{M_{\mathrm{J}}}\right)^{5/3}
\left(\frac{R_{\mathrm{J}}}{R}\right)^{3},
\label{eq:nup}
\end{equation}

Here $\omega$ ($\omega_{\mathrm{J}}$) is the planetary (Jovian) rotation rate,
$M$ ($M_{\mathrm{J}}$) the planetary (Jovian) mass,
$R$ ($R_{\mathrm{J}}$) the planetary (Jovian) radius,
$d$ the Earth--star distance, and $a$ the orbital semi-major axis.
In the fiducial model, close-in planets ($a < 0.1\,\mathrm{au}$) are assumed tidally locked ($\omega=2\pi/P_{\mathrm{orb}}$). More distant planets use Jupiter's spin rate ($\omega=\omega_{\mathrm{J}}$).
The radio beaming solid angle is fixed at the Jovian reference value $\Omega=1.6\,\mathrm{sr}$.
The stellar ionized mass-loss rate $\dot{M}_{\mathrm{ion}}$ and terminal wind speed $v_{\infty}$
follow empirical relations of \citet{Griemeier2005a}, with
$n(a)=n_{0}(1+t/\tau)^{-1.86}(a/1\,\mathrm{au})^{-2}$,
$n_{0}=1.04\times10^{11}\,\mathrm{m}^{-3}$, $\tau=2.56\times10^{7}\,\mathrm{yr}$,
and $v_{\infty}=0.75\,v_{\mathrm{esc}}$.

Rotation enters both $S_{\nu}$ and $\nu_{p}$. Close-in giants may not be fully synchronized, so the tidal-locking prescription for $a<0.1$\,au is an assumption rather than a measured spin. Yet spin is also a dynamo control. The beaming solid angle is fixed at the Jovian reference $\Omega=1.6$\,sr, whereas actual beaming depends on magnetospheric geometry and viewing angle. Stellar-wind mass loss and terminal speed follow the solar-calibrated density law of \citet{Griemeier2005a} and may be inaccurate for M dwarfs or young stars, precisely the hosts for which atmospheric erosion and habitability are most contested. That solar calibration is not used to rank white-dwarf hosts or catalog ages below $0.1$\,Gyr. Both cuts are defined in the sample selection above. Radio non-detections (or future detections) therefore constrain combinations of field strength, spin, and wind forcing, the same parameters that enter atmospheric-loss models, rather than any single factor in isolation.

\begin{figure}[!ht]
\centering
\includegraphics[width=0.85\textwidth]{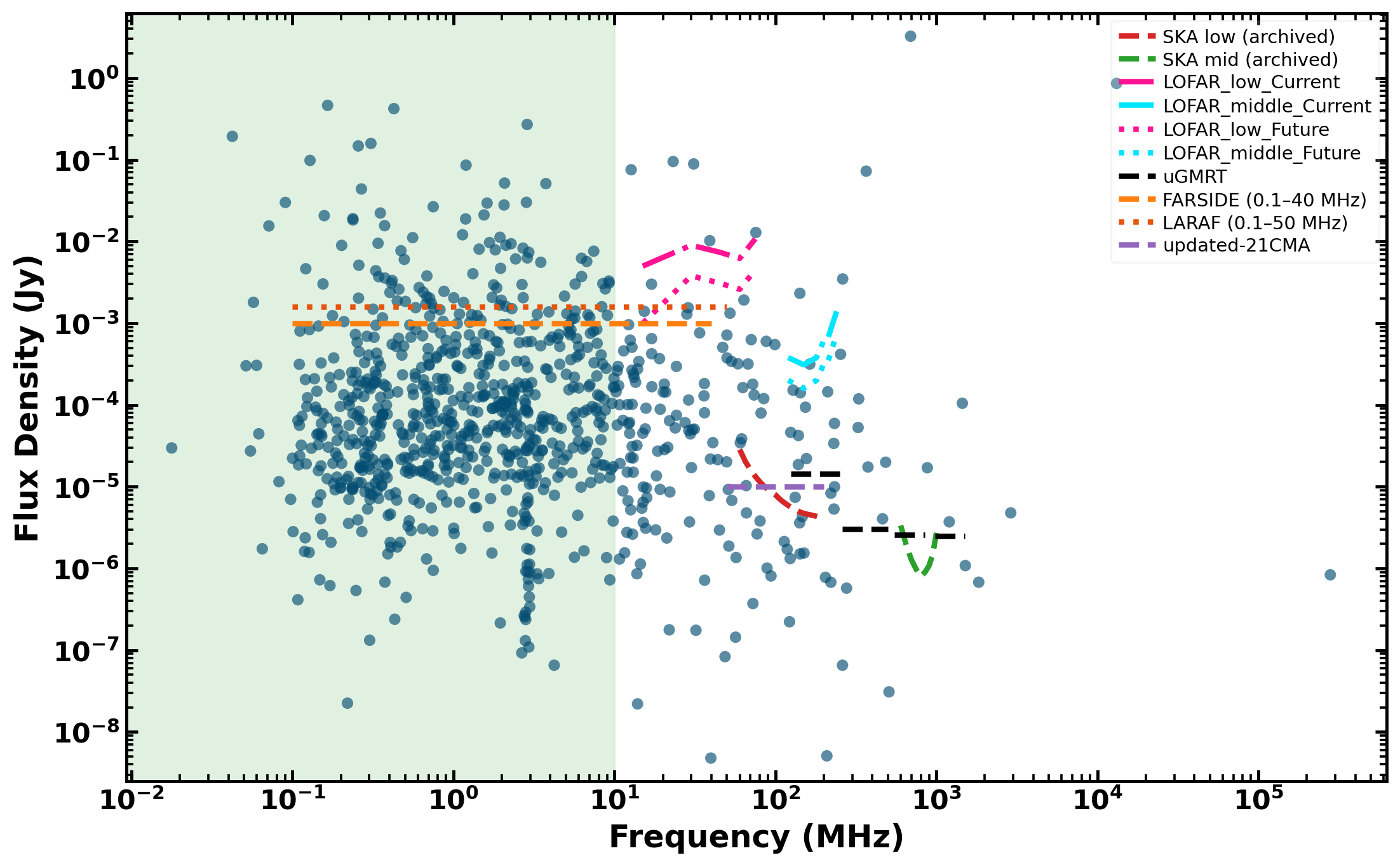}
\caption{Predicted characteristic frequency $\nu_{p}$ versus flux density $S_{\nu}$ for giant planets ($M_{p}>0.1\,M_{J}$; N = 898) predicted using the kinetic RBL. Colored curves are illustrative continuum sensitivity tracks (archived SKA low/mid tracks, LOFAR current/future, and uGMRT from archived facility tables; FARSIDE, LARAF, and updated-21CMA as constant band markers). The archived SKA curves are not the \citet{Braun2019} SKA1 bands in Table~\ref{tab:facilities}. They are not homogenized to common observing conditions and are not used for Figure~\ref{fig:inband}. The shaded region marks the ionospheric cutoff.}
\label{fig:detectability}
\end{figure}

\begin{figure}[!h]
\centering
\includegraphics[width=1\textwidth]{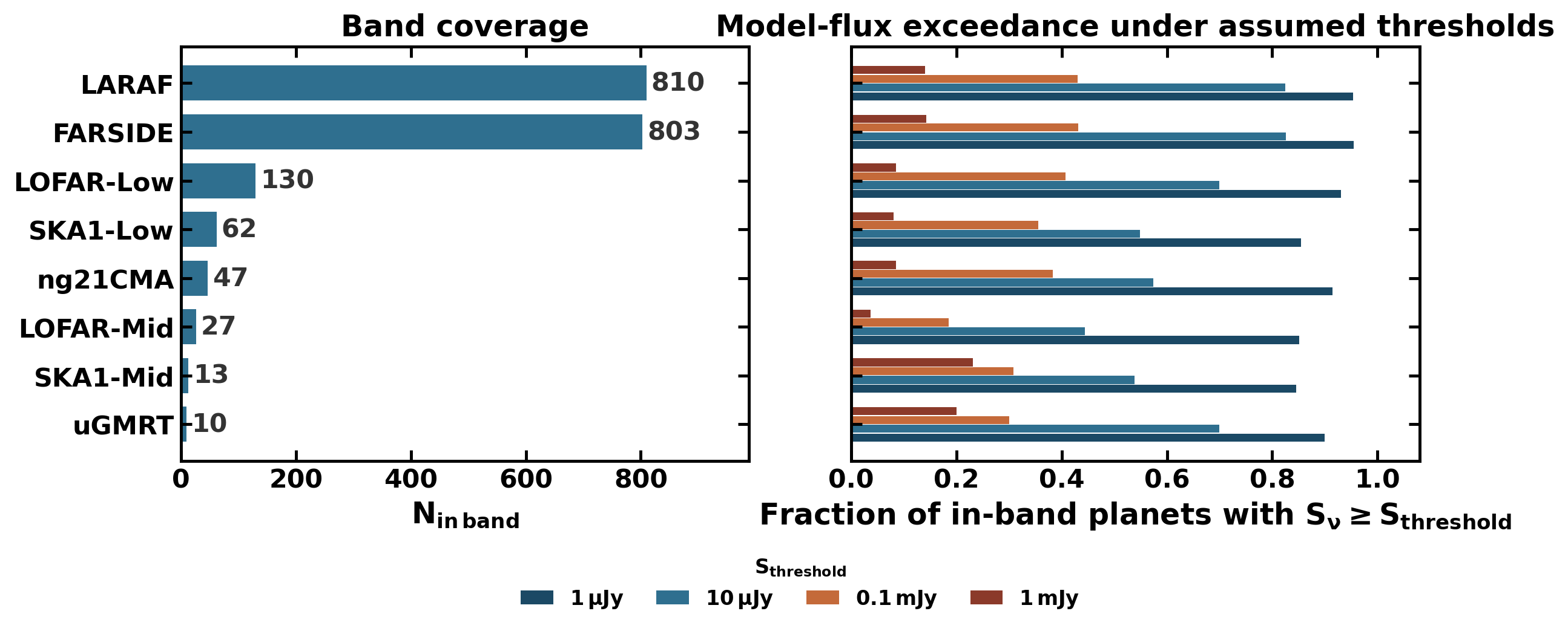}
\caption[{In-band facility comparison for the $N=898$ giant sample.}]{In-band facility comparison for the $N=898$ giant-planet mass-selected sample. Facilities are ordered by $N_{\mathrm{in\,band}}$. Left: frequency-window coverage $N_{\mathrm{in\,band}}$ using the bands in Table~\ref{tab:facilities}. $N_{\mathrm{in\,band}}$ counts planets whose characteristic frequency falls in the band. It does not count planets whose emission spectrum intersects the band. Right: fraction of in-band planets (denominator $N_{\mathrm{in\,band}}$, not the full sample of 898) with model $S_{\nu}\geq S_{\mathrm{threshold}}$ for unified assumed thresholds $1\,\mu\mathrm{Jy}$, $10\,\mu\mathrm{Jy}$, $0.1\,\mathrm{mJy}$, and $1\,\mathrm{mJy}$. These are kinetic-RBL model-flux exceedances under assumed continuum thresholds; they are not ECM detection probabilities and do not include beaming, duty cycle, propagation losses, or real-array sensitivity.\label{fig:inband}}
\end{figure}

Frequency-dependent curves in Figure~\ref{fig:detectability} are illustrative continuum sensitivity tracks for visual context only: archived SKA low/mid tracks, LOFAR LBA/HBA current and future tracks from archived LOFAR sensitivity tables, and uGMRT band floors from archived uGMRT sensitivity tables; the FARSIDE, LARAF, and updated-21CMA traces are constant illustrative levels over the nominal bands in Table~\ref{tab:facilities}. The archived SKA curves are not the \citet{Braun2019} SKA1 performance curves. Those curves are not homogenized to a common integration time, bandwidth, or detection significance, and they are not used for the Figure~\ref{fig:inband} statistics. Facility comparisons in Figure~\ref{fig:inband} use only the frequency bands in Table~\ref{tab:facilities}.

\begin{table}[!htbp]
\centering
\caption{Representative facility bands used for the Figure~\ref{fig:inband} coverage comparison. }
\label{tab:facilities}
\begin{tabular}{llp{6.8cm}}
\toprule
Facility & Band (MHz) & Reference \\
\midrule
FARSIDE & 0.1--40 & \citet{Burns2019} \\
LARAF & 0.1--50 & \citet{Chen2024} \\
LOFAR-Low & 10--90 & \citet{vanHaarlem2013} \\
LOFAR-Mid & 110--250 & \citet{vanHaarlem2013} \\
ng21CMA & 50--200 & \citet{Gu2026} \\
SKA1-Low & 50--350 & \citet{Braun2019} \\
SKA1-Mid & 350--15000 & \citet{Braun2019} \\
uGMRT & 300--1400 & \citet{Gupta2017} \\
\bottomrule
\end{tabular}
\end{table}

FARSIDE and LARAF are entered separately at the windows of the cited concepts, 0.1--40\,MHz and 0.1--50\,MHz. A single Moon-based window of 0.01--10\,MHz is not used. SKA1-Low and SKA1-Mid are the Phase~1 bands in \citet{Braun2019}. That reference is an SKA1 performance report and is not cited here as an SKA2 specification.

Figure~\ref{fig:inband} (left) reports \(N_{\mathrm{in\,band}}\), the number of giants in the \(N=898\) sample whose predicted characteristic frequency \(\nu_{p}\) falls in each facility band. The count uses one characteristic frequency per planet. It is not the number of planets whose emitted spectrum intersects the band. Real electron-cyclotron maser emission has a finite spectral range, so a characteristic frequency inside a facility window is a simplified coverage indicator and does not by itself establish that the radiation is observable. Figure~\ref{fig:inband} (right) reports the fraction of those in-band planets with model \(S_{\nu}\geq S_{\mathrm{threshold}}\), for unified assumed continuum thresholds \(S_{\mathrm{threshold}}\in\{1\,\mu\mathrm{Jy},\,10\,\mu\mathrm{Jy},\,0.1\,\mathrm{mJy},\,1\,\mathrm{mJy}\}\) applied identically to every facility. The denominator is therefore \(N_{\mathrm{in\,band}}\) for that facility and not the full sample of 898. These fractions are kinetic-RBL model-flux exceedance rates under assumed thresholds. They are not ECM detection probabilities and do not encode beaming, duty cycle, propagation losses, polarization filtering, RFI, confusion, imaging dynamic range, or real-array sensitivity.

Because many close-in giants may not be fully synchronized\, and because spin enters the magnetic-field scaling through $\nu_{p}\propto\omega$, we recompute $\nu_{p}$ under a fixed-rotation alternative with $\omega=\omega_{\rm J}$ for every planet. Separately, we rescale $\Omega$ by a factor of 3, $\dot{M}_{\rm ion}$ by a factor of 10, and an overall radio-efficiency factor by a factor of 10. The rotation experiment changes the frequency distribution, whereas the rescaling experiments change only the absolute $S_{\nu}$. Uniform multiplicative rescalings leave relative Top-20 ranks unchanged by construction; they therefore probe absolute flux scale rather than ranking robustness.

Target-dependent mass uncertainty is tested in two distinct ways (Section~\ref{sec:results}). (i)~Sample selection: recompute population statistics and the Top-20 on the catalog non-upper-limit mass subset alone ($N=786$), which changes sample membership. The overlap quoted for Table~\ref{tab:top20} removes the white-dwarf hosts, the catalog ages below $0.1$\,Gyr, and Kepler-415\,b from both lists. (ii)~Mass uncertainty: leave the fiducial sample membership fixed at $N=898$, hold case~(i) catalog masses fixed, and apply $N_{\mathrm{MC}}=1000$ Monte Carlo draws in which each of the 112 Chen–Kipping-estimated masses is independently perturbed as $\log_{10}M\leftarrow\log_{10}M+\delta$ with $\delta\sim\mathcal{N}(0,0.3\,\mathrm{dex})$. Both $S_{\nu}\propto M^{-0.33}$ and $\nu_{p}\propto M^{5/3}$ are updated in each draw (rotation and radius held fixed for each planet). We distinguish flux-ranking robustness (Top-20 overlap on $S_{\nu}$) from frequency-window robustness (the sample fraction with $\nu_{p}<10$\,MHz), reporting both for the fixed sample and after re-applying $M_{p}>0.1\,M_{J}$ within each draw (objects that fall below the cut are dropped from that realization).

Table~\ref{tab:searches} compiles published searches for exoplanetary radio emission. The limits span microjanskys to janskys and document both improving facility reach and the continuing absence of confirmed magnetospheric detections. This is the observational backdrop against which the population forecast must be judged.

\begin{table}[!hb]
\centering
\small
\caption{Published searches for exoplanetary radio emission. An entry is included only when the cited paper reports a telescope, an observing frequency or band, and a quantitative sensitivity or upper limit. The values are heterogeneous: they may be noise levels, upper limits, or ranges from different observing setups, and they are not normalized to a common integration time or detection statistic.}
\label{tab:searches}
\begin{tabular}{lp{2.5cm}p{2.7cm}p{5.4cm}}
\toprule
Telescope & Frequency & Sensitivity & Reference \\
\midrule
UTR-2 & $\sim$10--35 MHz & $\sim$3--6\,Jy & \citet{Zarka1997} \\
UTR-2 & 18--32 MHz & $\sim$1\,Jy & \citet{Ryabov2004} \\
\midrule
LOFAR & 10--90 MHz & $\sim$50 mJy & \citet{Turner2019} \\
LOFAR & 10--90 MHz & $\sim$50--73 mJy & \citet{Turner2021} \\
LOFAR & 30--78 MHz & $\sim$1--7 mJy & \citet{Gasperin2020} \\
LOFAR & 15--39 MHz & 165 mJy & \citet{Turner2024} \\
\midrule
MWA & 154 MHz & 3.4--49.9 mJy & \citet{Murphy2015} \\
MWA & 154 MHz & 4--235 mJy & \citet{Lynch2017} \\
\midrule
GMRT & $\sim$150 MHz & $\sim$2 mJy & \citet{Majid2006} \\
GMRT & 153 MHz & 1.2 mJy & \citet{Hallinan2013} \\
GMRT & 244 MHz & 2 mJy & \citet{Lecavelier2009} \\
GMRT & 614 MHz & 0.16 mJy & \citet{Lecavelier2009} \\
\midrule
GBT & 307--347 MHz & 81 mJy & \citet{Smith2009} \\
\midrule
VLA & 74 MHz & $\sim$50 mJy & \citet{Bastian2000} \\
VLA & 333 MHz & 1--10 mJy & \citet{Bastian2000} \\
VLA & 1465 MHz & 0.02--0.07 mJy & \citet{Bastian2000} \\
\midrule
Mizusawa & 8600 MHz & $\sim$1--10\,Jy & \citet{Shiratori2006} \\
\bottomrule
\end{tabular}
\end{table}

\section{Results}
\label{sec:results}

\begin{figure}[!h]
\centering
\includegraphics[width=0.9\textwidth]{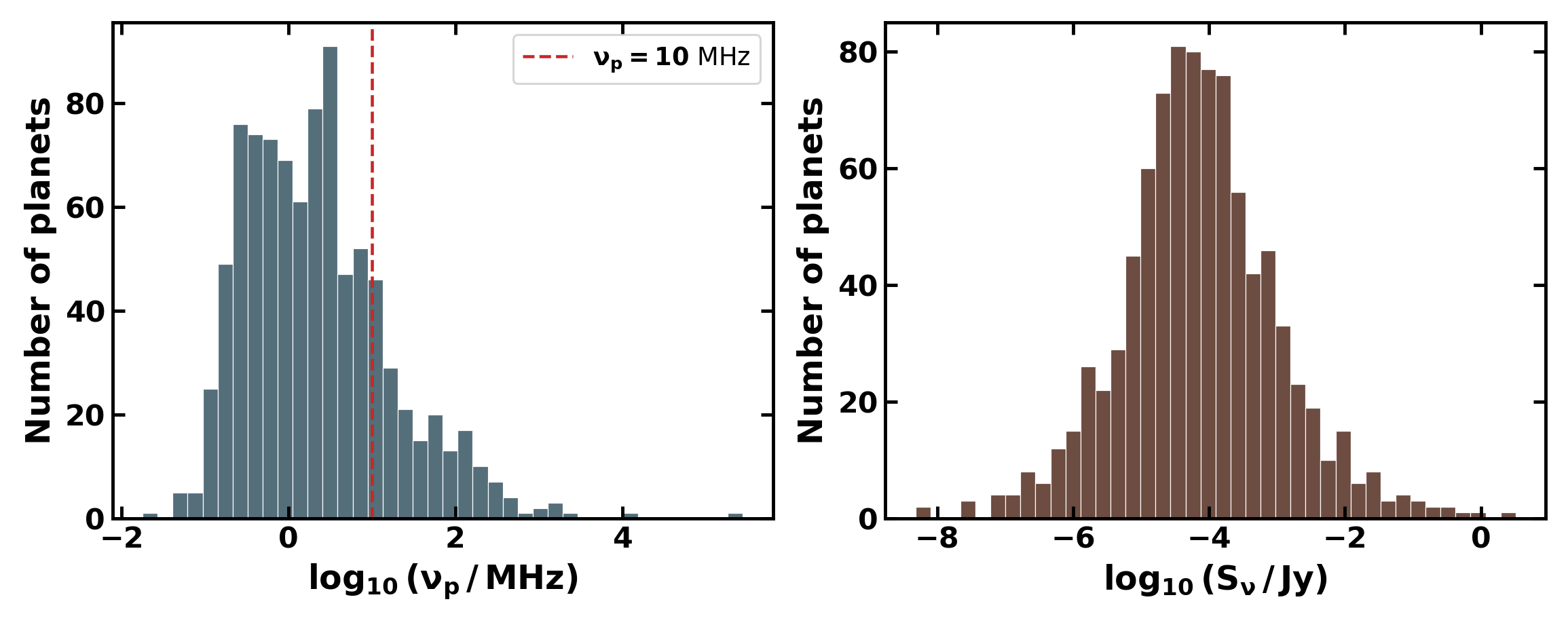}
\caption{The distributions of $\nu_p$ and $S_\nu$. The low-frequency peak places most objects below the ionospheric cutoff, a frequency regime that must be accessible for radio detections to constrain magnetic shielding of planetary atmospheres.}
\label{fig:histograms}
\end{figure}

Our analysis predicts the dominant frequency range of magnetospheric radio emission for the giant-planet mass-selected sample (Figures~\ref{fig:detectability} and~\ref{fig:histograms}). For the mass-selected sample (N = 898), the median characteristic frequency is 1.82\,MHz (16--84th percentile 0.30--13.7\,MHz), and 80.1\% of the sample lies below 10\,MHz. Median $S_{\nu}$ is only $6.4\times10^{-5}$\,Jy. Figure~\ref{fig:histograms} shows the corresponding $\nu_p$ and $S_\nu$ histograms for the same sample. Both distributions are strongly skewed: $\nu_p$ piles up below $\sim$10\,MHz, and $S_\nu$ remains faint for most objects. Hot, warm, and cold giants ($a<0.1$, $0.1\leq a<1$, and $a\geq1$\,au) number 722, 166, and 10 (Figure~\ref{fig:by_class}), so the low-frequency pile-up is carried by the close-in majority, precisely the planets whose magnetospheres, if present, mediate the strongest star--planet coupling. Its numerical coincidence with the terrestrial ionospheric cutoff is therefore accidental. The ionosphere blocks ground-based access near that frequency, but it does not create the RBL concentration. In habitability terms, the magnetospheres one most wants to detect around irradiated giants are predicted to radiate largely at frequencies inaccessible from the ground.

\begin{figure}[!ht]
\centering
\includegraphics[width=0.85\textwidth]{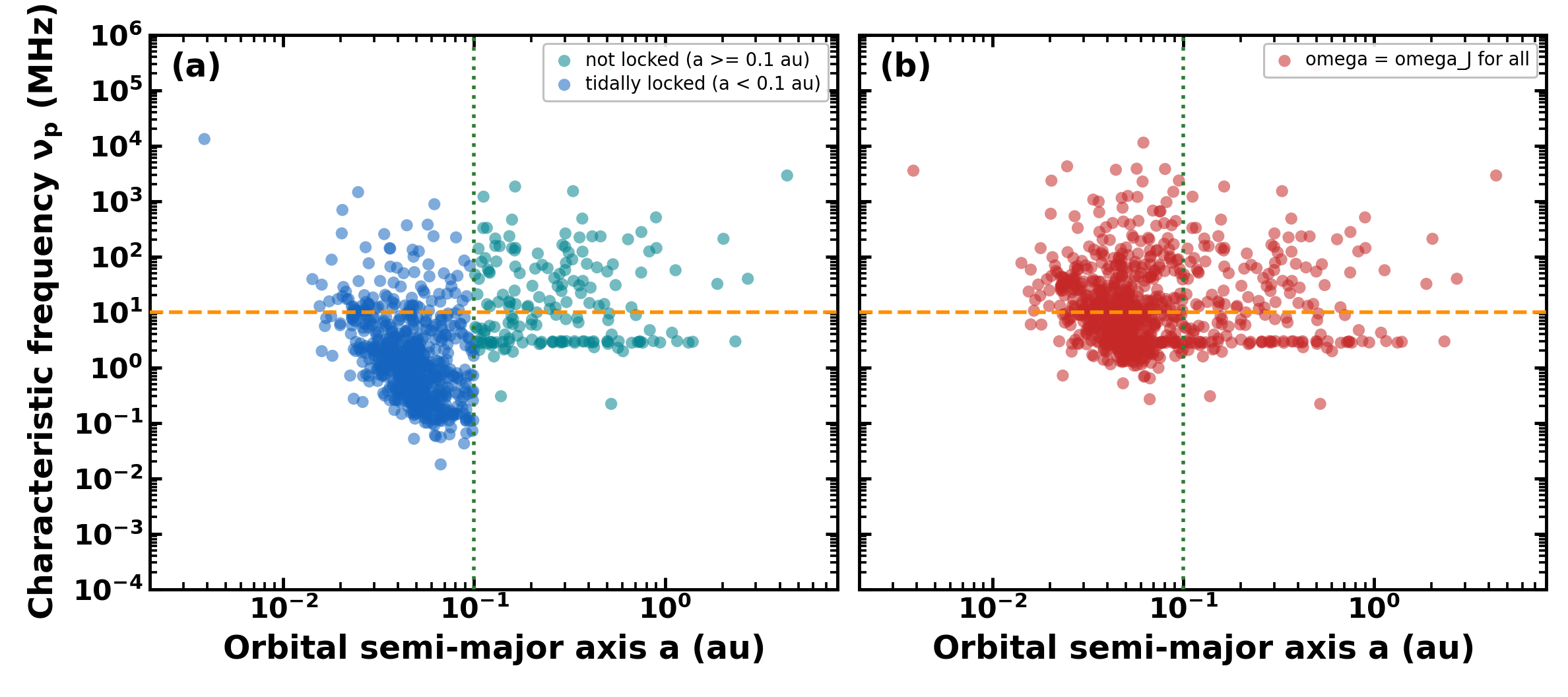}
\caption{The dependence of magnetospheric cyclotron frequency on adopted spin assumptions. Characteristic frequency is plotted against orbital separation for the giant planets. (a) Fiducial model. Planets with $a<0.1$\,au are tidally locked ($\omega=2\pi/P_{\mathrm{orb}}$; blue; N=722), while planets at larger separations adopt $\omega=\omega_{\mathrm{J}}$ (teal; N=176). (b) Fixed $\omega=\omega_{\mathrm{J}}$ is imposed for all planets. Dotted green line marks $a=0.1$\,au. Dashed orange line marks $\nu_p=10$\,MHz. In panel (a), the low-frequency pile-up arises primarily among tidally locked close-in giants. Removing the tidal-locking assumption in panel (b) increases $\nu_p$ at small orbital separations.}
\label{fig:tidal_lock}
\end{figure}

That interpretation is tested by changing only the rotation law (Figure~\ref{fig:tidal_lock}). Mapping $\nu_{p}$ against orbital separation shows that the low-frequency pile-up is localized to $a<0.1$\,au, where the fiducial model imposes tidal locking ($\omega=2\pi/P_{\mathrm{orb}}$) and therefore slower spins, the same close-in regime where stellar-wind forcing most strongly threatens atmospheric retention. Assigning $\omega=\omega_J$ to every planet raises the characteristic frequencies of most close-in planets and increases the sample median $\nu_p$ from 1.82 to 8.95 MHz, while reducing the fraction below 10 MHz from 80.1\% to 52.7\%. The $\sim$10\,MHz concentration is therefore spin-assumption dependent. Synchronization is simultaneously a dynamo control, a habitability-relevant boundary condition for irradiated giants, and a factor determining which frequency window can probe the field. By contrast, rescaling $\Omega$, $\dot{M}_{\mathrm{ion}}$, and radio efficiency changes absolute $S_{\nu}$ but not $\nu_{p}$. Absolute detectability remains uncertain. Relative priority among the brightest giants is more stable.

\begin{figure}[!ht]
\centering
\includegraphics[width=0.85\textwidth]{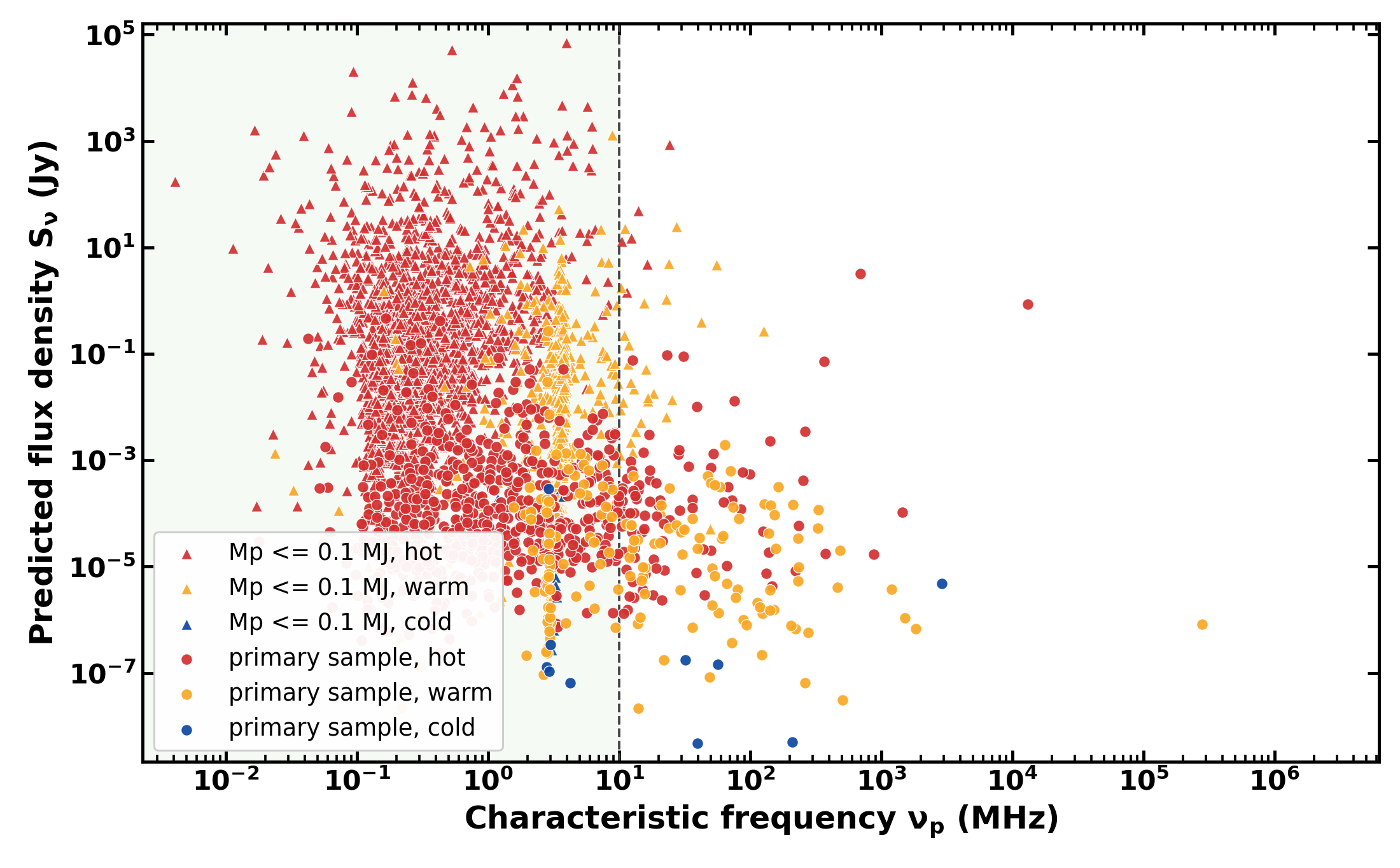}
\caption{Kinetic RBL predictions for the full exoplanet sample ($N=4428$). Color denotes orbital class (2784 hot, 1624 warm, 20 cold). Circles indicate giants ($M_{p}>0.1\,M_\mathrm{J}$; $N=898$: 722 hot, 166 warm, 10 cold). Triangles denote objects with $M_{p}\leq0.1\,M_\mathrm{J}$ ($N=3530$). The dashed line marks $\nu_p=10$\,MHz.}
\label{fig:by_class}
\end{figure}

Figure~\ref{fig:detectability} places the $(\nu_{p},S_{\nu})$ predictions against facility bands, and Figure~\ref{fig:inband} summarizes band coverage and assumed-threshold exceedance for the same $N=898$ catalog. Most of the mass-selected giant sample falls below $\sim$10\,MHz, so the magnetospheres that mediate the strongest wind--atmosphere coupling require lunar- or space-based low-frequency observations. Left-panel $N_{\mathrm{in\,band}}$ counts are FARSIDE $803$, LARAF $810$, LOFAR-Low $130$, SKA1-Low $62$, ng21CMA $47$, LOFAR-Mid $27$, SKA1-Mid $13$, and uGMRT $10$. FARSIDE and LARAF start at 0.1\,MHz: 12 giants have $\nu_{p}<0.1$\,MHz and fall outside both lunar windows, while part of each lunar count lies above 10\,MHz. The right panel shows how the fraction of those in-band planets with model $S_{\nu}\geq S_{\mathrm{threshold}}$ declines as the assumed threshold is raised from $1\,\mu\mathrm{Jy}$ to $1\,\mathrm{mJy}$ (e.g.\ $\sim$0.85 for SKA1-Low at $1\,\mu\mathrm{Jy}$ means $\sim$85\% of the 62 in-band targets, not 85\% of all 898). Those fractions are kinetic-RBL flux exceedances under unified continuum assumptions, not ECM detection probabilities or campaign yields. The absolute-$S_{\nu}$ ranking used for follow-up priority is Table~\ref{tab:top20} (e.g.\ TOI-1268\,b, TOI-3884\,b, TOI-1853\,b, and TOI-332\,b). The ten rows in Table~\ref{tab:rank_exclude} are not in that list. Because $S_{\nu}\propto(R/R_{\mathrm{J}})^{-3}$, small-radius entries can rise in the list. The catalog non-upper-limit comparison ($N=786$; NEA limit flag unset, not a literature-confirmed mass sample) changes sample composition: population medians remain similar (median $\nu_{p}=1.99$\,MHz; 77.2\% below 10\,MHz). On the raw $N=898$ order the Top-20 overlap is $15/20$, because five Chen \& Kipping masses leave. Those five are among the young hosts removed from the priority list, so the overlap of Table~\ref{tab:top20} with the catalog non-upper-limit ranking after the same cuts ($N=781$) is $20/20$. The priority sample ($N=888$) has median $\nu_{p}=1.80$\,MHz and 80.2\% below 10\,MHz, and the $N=781$ subset has median $\nu_{p}=1.98$\,MHz and 77.5\% below 10\,MHz. Removing the ten rows changes FARSIDE from 803 to 797, LARAF from 810 to 804, SKA1-Mid from 13 to 10, and uGMRT from 10 to 8. Separately, the Monte Carlo mass-uncertainty test ($N_{\mathrm{MC}}=1000$; independent $0.3$\,dex Gaussian draws in $\log_{10}M$ for each of the 112 estimated masses, case~(i) catalog masses fixed) updates both $S_{\nu}\propto M^{-0.33}$ and $\nu_{p}\propto M^{5/3}$. Flux-ranking robustness on the priority order is a mean Top-20 overlap of $1.00$ in every realization, both with fixed membership and after re-applying $M_{p}>0.1\,M_{J}$ (on average $\sim$22 estimated-mass objects fall below the cut per realization). None of the priority Top-20 is a Chen \& Kipping mass, and none of the perturbed masses enters that list. This overlap is numerical stability under that mass noise. It does not test the stellar-wind law, host age, spin state, or beaming direction. The same Monte Carlo on the \(N=898\) order, which still places the excluded rows first, gives $0.98$ ($16$--$84$th percentile $0.95$--$1.00$) and $0.94$ ($0.90$--$1.00$). Frequency-window robustness: the fiducial fraction below 10\,MHz ($80.1\%$) becomes $79.4\%$ (16--84th percentile $79.2$--$79.7\%$) with fixed membership and $78.9\%$ ($78.7$--$79.2\%$) after the mass cut, so the sub-ionospheric pile-up is stable even though individual $\nu_{p}$ values scale steeply with mass. The $20/20$ overlap is the overlap of Table~\ref{tab:top20}. The $15/20$ figure is only the raw \(N=898\) order against its catalog non-upper-limit Top-20. These overlaps remain complementary to the Monte Carlo (sample selection versus mass noise), and neither should be confused with frequency-window stability, which is reported separately above. Such ranks are kinetic priorities for magnetospheric follow-up, not automatic claims that every catalog radius hosts a Jupiter-analogue field, and thus not direct assessments of habitability.

\begin{figure*}[!ht]
\centering
\includegraphics[width=0.85\textwidth]{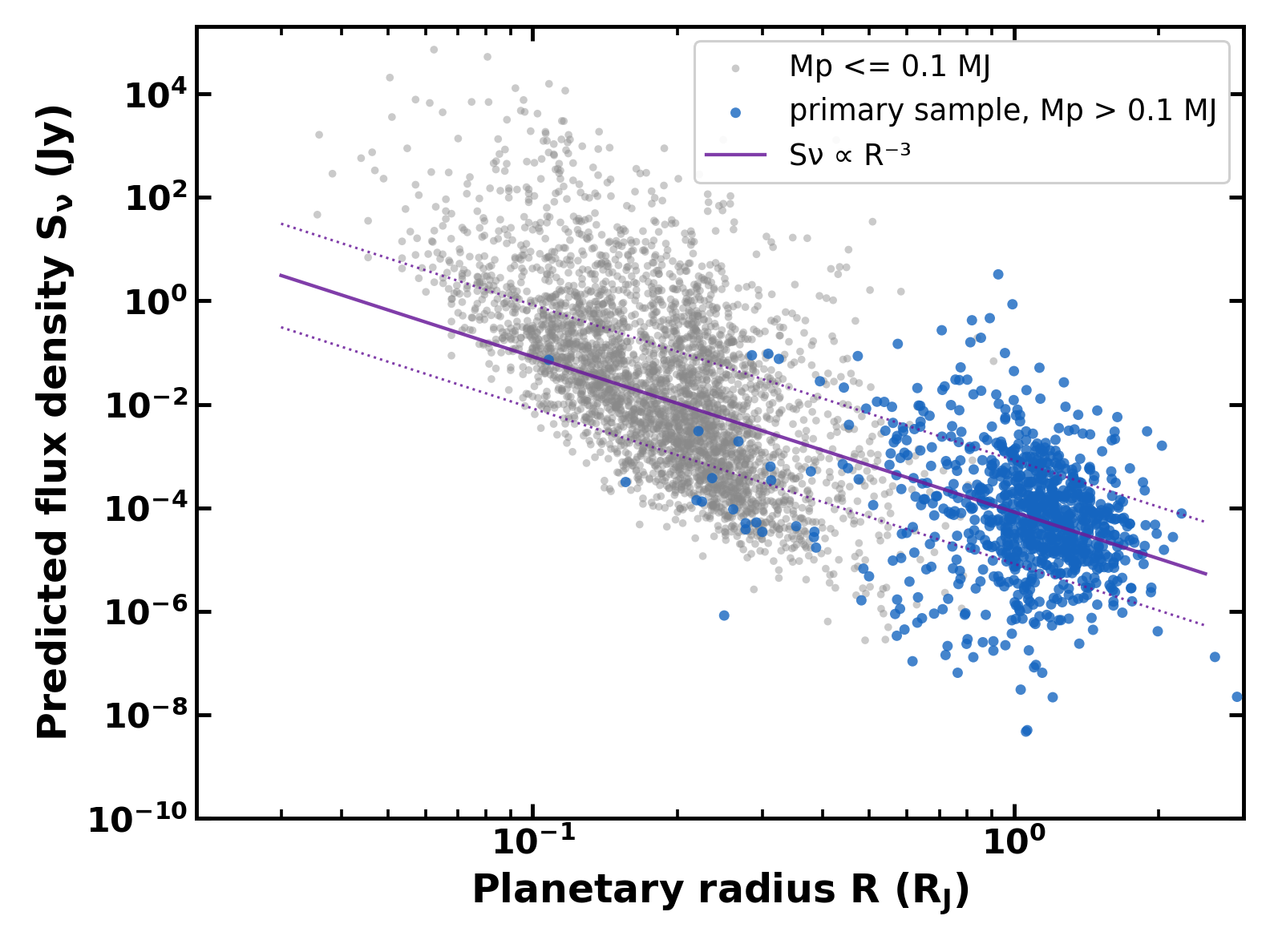}
\caption{Kinetic RBL predictions of $S_\nu$ versus planetary radius for the full Archive working sample ($N=4428$). Blue points denote the giant-planet mass-selected sample} ($M_p>0.1\,M_\mathrm{J}$; $N=898$). Grey points denote objects with $M_p\leq0.1\,M_\mathrm{J}$ ($N=3530$), below that working cut, where a Jupiter-analogue magnetic field, and therefore magnetic atmospheric shielding, is least well constrained. Purple reference tracks follow $S_{\nu}\propto R^{-3}$ (solid curve normalized near $R\sim R_\mathrm{J}$; dotted curves scaled by factors of 0.1 and 10), highlighting the radial scaling in Equation~\eqref{eq:Snu}.
\label{fig:extrapolation}
\end{figure*}

Figure~\ref{fig:by_class} places the giants against the full Archive working sample (N = 4428) under the same kinetic formulae. Color encodes orbital class and marker encodes the mass cut. Filled circles are the primary giant-planet mass-selected sample ($M_{p}>0.1\,M_{J}$; N = 898), while triangles are $M_{p}\leq0.1\,M_{J}$ objects below that working cut. The lower-mass cloud occupies a similar low-frequency region, but that is exactly where a Jupiter-analogue magnetosphere, and thus a direct RBL reading as a magnetic or habitability diagnostic, is least secure. Population forecasts and survey strategy therefore focus on the mass-selected giant sample first. Figure~\ref{fig:extrapolation} shows $S_{\nu}$ versus planetary radius for the same mass cut, with $S_{\nu}\propto R^{-3}$ reference tracks that expose the size scaling built into Equation~\eqref{eq:Snu}.

\begin{table}[!ht]
\centering
\footnotesize
\setlength{\tabcolsep}{3.2pt}
\caption{Rows removed from the priority ranking. They remain in the $N=898$ population totals. Asterisks mark Chen \& Kipping masses. Ages are catalog values. ZTF~J1828+2308\,b has no catalog age. Kepler-415\,b is omitted for an unreliable catalog $M$--$R$ pair (see Discussion), not for host type or age.}
\label{tab:rank_exclude}
\begin{tabular}{l l r r}
\toprule
Planet & Reason & $S_{\nu}$ (Jy) & $\nu_{p}$ (MHz) \\
\midrule
WD 1856+534 b & white-dwarf host; $13.8\,M_{J}$ is a 95\% upper limit & 3.25 & 691 \\
ZTF J1828+2308 b & white-dwarf host; brown dwarf & 0.859 & $1.32\times10^{4}$ \\
HIP 67522 b$^{\ast}$ & 17\,Myr & $0.463$ & 0.165 \\
TOI-837 b & 35\,Myr & $0.422$ & 0.427 \\
HIP 67522 c$^{\ast}$ & 17\,Myr & $0.270$ & 2.87 \\
TOI-1227 b$^{\ast}$ & 11\,Myr & $0.193$ & 0.0423 \\
IRAS 04125+2902 b$^{\ast}$ & 3.3\,Myr & $9.81\times10^{-2}$ & 0.129 \\
TOI-6448 b$^{\ast}$ & 34\,Myr & $3.01\times10^{-2}$ & 2.84 \\
NGTS-33 b & 30\,Myr & $5.70\times10^{-3}$ & 6.73 \\
Kepler-415 b & extreme $M/R$; TTV mass highly uncertain & $7.25\times10^{-2}$ & 366 \\
\bottomrule
\end{tabular}
\end{table}

\begin{table}[!ht]
\centering
\footnotesize
\setlength{\tabcolsep}{3.2pt}
\caption{Priority Top-20 after white-dwarf hosts and catalog ages below $0.1$\,Gyr are removed and after omitting Kepler-415\,b ($N=888$). None of these masses is a Chen \& Kipping estimate. Eight of the twenty have a catalog age of exactly $4.6$\,Gyr, which is also the fill used when the age is missing, so that value is not read as a measured solar age. This is the ranking referred to in the text. The $N=898$ population totals still contain the ten excluded rows as numerical entries.}
\label{tab:top20}
\begin{tabular}{r l r r r r r r l}
\toprule
Rank & Planet & $S_{\nu}$ (Jy) & $\nu_{p}$ (MHz) & $d$ (pc) &
$a$ (au) & $M_{p}$ ($M_{\mathrm{J}}$) & $R_{p}$ ($R_{\mathrm{J}}$) & Class \\
\midrule
1 & TOI-1268 b & $1.58\times10^{-1}$ & 0.307 & 109.6 & 0.0711 & 0.303 & 0.812 & hot \\
2 & TOI-3884 b & $1.48\times10^{-1}$ & 0.256 & 43.14 & 0.0361 & 0.103 & 0.574 & hot \\
3 & TOI-1853 b & $9.52\times10^{-2}$ & 23.2 & 165.1 & 0.0213 & 0.23 & 0.309 & hot \\
4 & TOI-332 b & $8.90\times10^{-2}$ & 31.1 & 222.9 & 0.0159 & 0.18 & 0.285 & hot \\
5 & TOI-672 b & $8.59\times10^{-2}$ & 1.2 & 66.91 & 0.0376 & 0.16 & 0.474 & hot \\
6 & TOI-849 b & $7.56\times10^{-2}$ & 12.7 & 225.7 & 0.0155 & 0.132 & 0.325 & hot \\
7 & WASP-59 b & $5.19\times10^{-2}$ & 2.08 & 115.9 & 0.0697 & 0.863 & 0.775 & hot \\
8 & HD 189733 b & $5.11\times10^{-2}$ & 3.75 & 19.76 & 0.0308 & 1.13 & 1.13 & hot \\
9 & WASP-69 b & $4.40\times10^{-2}$ & 0.268 & 49.96 & 0.0453 & 0.26 & 1 & hot \\
10 & WASP-139 b & $3.00\times10^{-2}$ & 0.0904 & 212.5 & 0.062 & 0.117 & 0.8 & hot \\
11 & TOI-4860 b & $2.95\times10^{-2}$ & 1.62 & 80.14 & 0.0181 & 0.273 & 0.77 & hot \\
12 & K2-55 b & $2.80\times10^{-2}$ & 2.06 & 159.5 & 0.0347 & 0.138 & 0.395 & hot \\
13 & WASP-52 b & $2.66\times10^{-2}$ & 0.749 & 174.8 & 0.0272 & 0.46 & 1.27 & hot \\
14 & TOI-1775 b & $2.23\times10^{-2}$ & 0.351 & 149.2 & 0.08 & 0.302 & 0.718 & hot \\
15 & TOI-1288 b & $2.11\times10^{-2}$ & 1.55 & 114.9 & 0.0374 & 0.139 & 0.443 & hot \\
16 & WASP-166 b & $2.07\times10^{-2}$ & 0.158 & 114.2 & 0.0641 & 0.101 & 0.63 & hot \\
17 & TOI-3984 A b & $1.90\times10^{-2}$ & 0.237 & 108.9 & 0.041 & 0.14 & 0.71 & hot \\
18 & Qatar-6 b & $1.89\times10^{-2}$ & 1.19 & 100.8 & 0.0423 & 0.668 & 1.06 & hot \\
19 & TOI-6894 b & $1.83\times10^{-2}$ & 0.238 & 73.81 & 0.026 & 0.168 & 0.855 & hot \\
20 & TOI-622 b & $1.56\times10^{-2}$ & 0.374 & 123.3 & 0.0708 & 0.303 & 0.824 & hot \\
\bottomrule
\end{tabular}
\end{table}

\section{Discussion}
\label{sec:discussion}
The primary sample is defined by a giant-planet mass cut ($M_{p}>0.1\,M_{J}$), chosen because a giant-planet RBL extrapolation is considered more plausible there than for lower-mass worlds; it is not a Solar System--validated applicability boundary for every object above that mass. Upper-limit-only catalog masses are excluded from the $N=898$ sample: a numeric limit is neither treated as a measured mass nor replaced by a Chen \& Kipping estimate. Population-level conclusions are stable under that cut, under the catalog non-upper-limit comparison ($N=786$), and after the priority cuts ($N=888$: 80.2\% below 10\,MHz; $N=781$: 77.5\% below 10\,MHz). The Monte Carlo mass-uncertainty test ($0.3$\,dex; $N_{\mathrm{MC}}=1000$) updates both $S_{\nu}$ and $\nu_{p}$. On the priority order the flux-ranking Top-20 overlap is $1.00$/$1.00$ under that mass perturbation, and the fraction below 10\,MHz stays near the fiducial priority-sample $80.2\%$ (the $N=898$ population figure remains $80.1\%$). The overlap does not measure stability against the stellar-wind law, host age, spin state, or beaming direction. The \(N=898\) order that still ranks the excluded rows first gives $0.98$/$0.94$. The $20/20$ overlap of Table~\ref{tab:top20} with the $N=781$ ranking reflects the removal of the Chen \& Kipping masses that had produced the raw $15/20$, not a conflict with either Monte Carlo metric.

An additional limitation concerns the treatment of missing stellar ages. We adopt a fiducial age of 4.6 Gyr when no catalog age is available, and several high-ranked targets are assigned this value. Because the adopted stellar-wind density prescription depends strongly on age, uncertainties in these assumed ages can substantially affect the predicted flux densities and relative rankings. Our mass-perturbation experiment does not propagate this source of uncertainty. The priority list should therefore be interpreted as a model-dependent reference ranking rather than a statistically robust ordering of observational targets.

We adopt the characteristic-frequency scaling $\nu_{p}\propto\omega M^{5/3}(R_{\mathrm{J}}/R)^{3}$ in Equation~\eqref{eq:nup}, following \citet{Li2023} and \citet{Narang2024} and the surface-field form in \citet{Farrell1999}. The frequency is the surface electron-cyclotron frequency. For a dipole, $B\propto\mu/R^{3}$, and therefore $\nu_{p}\propto\mu/R^{3}$. The radius exponent is exactly $-3$ under the assumption that, at fixed mass and rotation rate, the magnetic moment $\mu$ has no additional explicit radius dependence, so that $\omega$ and $M^{5/3}$ are carried entirely by $\mu$. The prefactor $23.5$\,MHz is the Jupiter normalization coefficient of this model. It is not a cutoff frequency common to every planet. This radius dependence is model-specific and should not be regarded as a universal planetary dynamo scaling. The mass-selected sample also includes compact, high-density planets whose radii are substantially smaller than those of typical gas giants. Although these objects satisfy the adopted mass threshold, their magnetospheric properties need not resemble those of Jupiter. In particular, the strong inverse-radius dependence of the kinetic-RBL flux prescription can produce large predicted flux densities for compact planets. Such predictions should be treated as model extrapolations rather than direct evidence of unusually strong or readily detectable planetary radio emission. Two catalog examples illustrate the point. Kepler-415\,b enters the NEA default table with $M_p = 0.377^{+0.321}_{-0.275}\,M_J$ and $R_{p}=0.108\,R_{J}$ from a TTV mass solution whose uncertainties are comparable to the mass itself. These parameters do not establish a robust measurement of an exceptionally high bulk density. KOI-2513.01 \citep{Canas2023} has $M_{p}=23\,M_{J}$ with $R_{p}=0.25\,R_{J}$ in the default table, but the published radius is essentially unconstrained ($R_{p}=0.25^{+2.0}_{-0.05}\,R_{J}$), so the extreme density is not a secure physical parameter. Because $S_{\nu}\propto R^{-3}$, neither object should be treated as an observational-priority target on the basis of the kinetic RBL alone. Kepler-415\,b is therefore removed from Table~\ref{tab:top20} and listed in Table~\ref{tab:rank_exclude}; KOI-2513.01 remains only in the appendix catalog with that caveat. Similar caution applies to other compact objects with unusually large catalog mass-to-radius ratios: uncertain masses or radii can make their inferred bulk densities unreliable, whereas even well-constrained parameters do not establish the applicability of a Jupiter-calibrated kinetic RBL. Given the strong dependence $S_\nu \propto R_p^{-3}$, their predicted flux densities should be treated as conditional model extrapolations rather than sufficient grounds for observational prioritization. The same caution applies to inflated hot Jupiters, high-mass brown-dwarf candidates, young giants, and systems with atypical hosts or dynamo geometries. This caveat does not justify retaining white-dwarf hosts or catalog ages below $0.1$\,Gyr in the priority ranks. Those host/age cuts, together with Kepler-415\,b, give the ten rows omitted from Table~\ref{tab:top20} and from the ordered catalog (Table~\ref{tab:top_targets}) and listed in Table~\ref{tab:rank_exclude}. The catalog mass of WD~1856+534\,b is a 95\% upper limit \citep{Vanderburg2020}, ZTF~J1828+2308\,b is a $\sim$19.5\,$M_{J}$ brown dwarf in the discovery paper \citep{Parsons2025}, and the solar-calibrated wind law does not apply to either host or to the seven hosts younger than $0.1$\,Gyr. Tidal synchronization for $a<0.1$\,au is likewise an assumption: relative to the tidally locked case, a spin rate faster than the orbital mean motion would raise $\nu_{p}$ for close-in planets and shrink the sub-ionospheric pile-up, whereas slower-than-synchronous rotation would lower $\nu_{p}$, consistent with the explicit dependence of \(\nu_p\) on the adopted rotation rate. The sub-10\,MHz concentration is therefore not rotation-independent: adopting Jupiter-like spins reduces the fraction below 10\,MHz from 80.1\% to 52.7\%. The observational accessibility of the giant-planet sample is itself model-dependent, which matters for the science case of space- or lunar-based low-frequency arrays. Stellar-wind forcing follows a solar-calibrated $\dot{M}_{\mathrm{ion}}$--$v_{\infty}$ prescription \citep{Griemeier2005a}; active M dwarfs and young stars can drive stronger, more variable winds, changing the absolute $S_{\nu}$ substantially without affecting $\nu_{p}$.

Even when $(\nu_{p},S_{\nu})$ fall inside a facility band, detectability is not guaranteed. Electron--cyclotron maser emission is beamed into a hollow cone, so geometric misalignment can hide an otherwise bright source. Propagation through the stellar wind and planetary ionosphere can quench or refract the radiation below the local plasma frequency, an effect outside Equations~\eqref{eq:Snu}--\eqref{eq:nup}. Stellar coronal bursts can dominate the same frequencies. Figure~\ref{fig:inband} therefore separates band coverage ($N_{\mathrm{in\,band}}$) from assumed-threshold model-flux exceedance among in-band planets; the right-panel fractions are not ECM detection probabilities and are not fractions of the full $N=898$ sample. Observational priority should combine high predicted $S_{\nu}$, accessible $\nu_{p}$, favorable geometry, and low stellar contamination, and should be followed by facility-specific sensitivity calculations in the appropriate observing mode.

Relative to prior population RBL applications such as \citet{MousaviSadr2025}, the incremental contribution here is limited to: (i)~a giant-planet mass-selected working sample with the exclusion of catalog mass upper limits; (ii)~explicit spin dependence of the $\sim$10\,MHz pile-up; and (iii)~facility-band coverage plus assumed-threshold model-flux exceedance (Figure~\ref{fig:inband}), rather than a claim of validated RBL detectability. Ranking and frequency-window tests constrain model robustness within those assumptions only. Historical non-detections (Table~\ref{tab:searches}) remain consistent with the faint median $S_{\nu}$ and with observing above the ionospheric cutoff for much of the sample.

\section{Conclusions}
\label{sec:conclusions}

For the giant-planet sample ($M_{p}>0.1\,M_{J}$; N = 898), the kinetic RBL predicts a median characteristic frequency $\nu_{p}=1.82$\,MHz (16--84th percentile $0.30$--$13.7$\,MHz), with $80.1\%$ of the sample below $10$\,MHz, and a median flux density $S_{\nu}=6.4\times10^{-5}$\,Jy. Hot, warm, and cold giants number $722$, $166$, and $10$, so the low-frequency concentration is carried by close-in planets (Figure~\ref{fig:by_class}).

Under the fiducial tidal-locking prescription, most giants lie below $\sim$10\,MHz, where only space- or lunar-based arrays open the window. That accessibility depends on the adopted rotation prescription: assigning Jupiter’s rotation rate to all planets reduces the fraction below 10\,MHz from 80.1\% to 52.7\%, so the science case for space- or lunar-based low-frequency arrays depends on the adopted spin prescription. Uniform rescaling of the beaming solid angle $\Omega$, stellar-wind mass-loss rate $\dot{M}_{\mathrm{ion}}$, and radio efficiency by factors of a few to ten mainly changes absolute $S_{\nu}$ and leaves relative ranks unchanged by construction. The catalog non-upper-limit cut and the priority cuts ($N=888$, 80.2\% below 10\,MHz; Top-20 overlap $20/20$ against the $N=781$ subset) probe sample selection. The $0.3$\,dex Monte Carlo updates both $S_{\nu}$ and $\nu_{p}$. On the priority order the mean Top-20 overlap is $1.00$/$1.00$ under that mass perturbation, and the $<$10\,MHz fraction stays near $80.2\%$ (the $N=898$ population figure remains $80.1\%$). That number is not a claim of stability under the stellar-wind law, host age, spin state, or beaming direction.

Figure~\ref{fig:inband} shows that facility relevance depends primarily on frequency coverage: most giants lie below $\sim$10\,MHz. FARSIDE (0.1--40\,MHz) and LARAF (0.1--50\,MHz) are the cited lunar concepts ($N_{\mathrm{in\,band}}=803$ and $810$). Ground-based bands recover fewer in-band sources (LOFAR-Low $130$; SKA1-Low $62$; ng21CMA $47$; LOFAR-Mid $27$; SKA1-Mid $13$; uGMRT $10$). Under unified assumed $S_{\mathrm{threshold}}$ values, the model-flux exceedance fraction among in-band planets falls as the threshold is raised from $1\,\mu\mathrm{Jy}$ to $1\,\mathrm{mJy}$; those fractions are not ECM detection probabilities, not fractions of all 898 giants, and not real-array yields. The priority ranking (Table~\ref{tab:top20}; e.g.\ TOI-1268\,b, TOI-3884\,b, TOI-1853\,b, TOI-332\,b) gives high-flux candidates under the adopted kinetic RBL model after white-dwarf hosts and catalog ages below $0.1$\,Gyr are removed , and after Kepler-415\,b is omitted for an unreliable $M$--$R$ pair. Those ten objects are listed in Table~\ref{tab:rank_exclude} and are not ranked. Even the objects in Table~\ref{tab:top20} are only conditionally recommended as follow-up targets, subject to frequency accessibility, instrumental sensitivity, emission beaming, and stellar contamination. They should not be regarded as automatic observational priorities. Historical non-detections (Table~\ref{tab:searches}) are consistent with the faint median flux and with observing above the ionospheric cutoff. Constraining the magnetospheres that may mediate atmospheric retention will require both deep LOFAR/SKA-class campaigns on the high-$\nu_{p}$ minority and, under the fiducial rotation assumption, low-frequency access from above the ionosphere for a substantial fraction of the sample.

\section*{Acknowledgements}

Hui Zhang acknowledges support from the 100 Talents Programme of the Chinese Academy of Sciences. Xiaolong Yang thanks the National Natural Science Foundation of China (12103076) and the Shanghai Sailing Program (21YF1455300) for support. Quan Guo acknowledges support from the Ministry of Science and Technology of China (grant No. 2020SKA0110200).

\section*{Appendix A. Primary-sample RBL target list}
\label{sec:appendix_targets}

Table~\ref{tab:top_targets} lists the flux-ordered giant targets after the white-dwarf hosts and the catalog ages below $0.1$\,Gyr are removed, and after Kepler-415\,b is omitted ($N=888$). The table is a supplementary list for magnetospheric follow-up aimed at calibrating the star--planet coupling relevant to atmospheric-loss and habitability assessments. Distances are system distances from the NASA Exoplanet Archive. Asterisks mark case (ii) mass estimates based on \cite{Chen2017}; upper-limit-only catalog masses are excluded from the sample entirely. Population figures in the main text still use $N=898$, which retains the ten excluded rows as numerical entries only. The same section reports the totals after those rows are removed ($N=888$ and the catalog non-upper-limit subset $N=781$). KOI-2513.01 remains in this appendix list: its default catalog mass and radius imply an extreme density, but the published radius is essentially unconstrained, so its kinetic-RBL prediction is not an observational priority.

\nolinenumbers 
\begin{longtable}{r l r r r r r r l}
\caption{Flux-ordered giants after white-dwarf hosts and catalog ages below $0.1$\,Gyr are removed and after omitting Kepler-415\,b ($N=888$). KOI-2513.01 is retained here only as a catalog entry; its extreme $M/R$ is not a secure priority target (Discussion).}
\label{tab:top_targets}\\
\toprule
Rank & Planet & $S_{\nu}$ (Jy) & $\nu_{p}$ (MHz) & $d$ (pc) &
$a$ (au) & $M_{p}$ ($M_{\mathrm{J}}$) & $R_{p}$ ($R_{\mathrm{J}}$) & Class \\
\midrule
\endfirsthead

\toprule
Rank & Planet & $S_{\nu}$ (Jy) & $\nu_{p}$ (MHz) & $d$ (pc) &
$a$ (au) & $M_{p}$ ($M_{\mathrm{J}}$) & $R_{p}$ ($R_{\mathrm{J}}$) & Class \\
\midrule
\endhead

\midrule
\multicolumn{9}{r}{Continued on next page} \\
\endfoot

\bottomrule
\endlastfoot
1 & TOI-1268 b & $1.58\times10^{-1}$ & 0.307 & 109.6 & 0.0711 & 0.303 & 0.812 & hot \\
2 & TOI-3884 b & $1.48\times10^{-1}$ & 0.256 & 43.14 & 0.0361 & 0.103 & 0.574 & hot \\
3 & TOI-1853 b & $9.52\times10^{-2}$ & 23.2 & 165.1 & 0.0213 & 0.23 & 0.309 & hot \\
4 & TOI-332 b & $8.90\times10^{-2}$ & 31.1 & 222.9 & 0.0159 & 0.18 & 0.285 & hot \\
5 & TOI-672 b & $8.59\times10^{-2}$ & 1.2 & 66.91 & 0.0376 & 0.16 & 0.474 & hot \\
6 & TOI-849 b & $7.56\times10^{-2}$ & 12.7 & 225.7 & 0.0155 & 0.132 & 0.325 & hot \\
7 & WASP-59 b & $5.19\times10^{-2}$ & 2.08 & 115.9 & 0.0697 & 0.863 & 0.775 & hot \\
8 & HD 189733 b & $5.11\times10^{-2}$ & 3.75 & 19.76 & 0.0308 & 1.13 & 1.13 & hot \\
\end{longtable}

\bibliography{zhao2026}

@ARTICLE{Ashtari2022,
       author = {{Ashtari}, Reza and {Sciola}, Anthony and {Turner}, Jake D. and others},
        title = "{Detecting Magnetospheric Radio Emission from Giant Exoplanets}",
      journal = {\apj},
         year = 2022,
        month = nov,
       volume = {939},
       number = {1},
        pages = {24},
          doi = {10.3847/1538-4357/ac92f5},
       adsurl = {https://ui.adsabs.harvard.edu/abs/2022ApJ...939...24A},
}

@ARTICLE{Bastian2000,
       author = {{Bastian}, T. S. and {Dulk}, G. A. and {Leblanc}, Y.},
        title = "{A Search for Radio Emission from Extrasolar Planets}",
      journal = {\apj},
         year = 2000,
        month = dec,
       volume = {545},
       number = {2},
        pages = {1058-1063},
          doi = {10.1086/317864},
       adsurl = {https://ui.adsabs.harvard.edu/abs/2000ApJ...545.1058B},
}

@ARTICLE{Brain2024,
       author = {{Brain}, David A. and {Kao}, Melodie M. and {O'Rourke}, Joseph G. and others},
        title = "{Exoplanet Magnetic Fields}",
      journal = {arXiv e-prints},
         year = 2024,
archivePrefix = {arXiv},
       eprint = {2404.15429},
 primaryClass = {astro-ph.EP},
       adsurl = {https://ui.adsabs.harvard.edu/abs/2024arXiv240415429B},
}

@ARTICLE{Burns2019,
       author = {{Burns}, Jack O. and {Hallinan}, Gregg and {Lux}, Jim and others},
        title = "{NASA Probe Study Report: Farside Array for Radio Science Investigations of the Dark ages and Exoplanets (FARSIDE)}",
      journal = {NASA Probe Study Report},
         year = 2019,
archivePrefix = {arXiv},
       eprint = {1911.08649},
 primaryClass = {astro-ph.IM},
       adsurl = {https://ui.adsabs.harvard.edu/abs/2019arXiv191108649B},
}

@ARTICLE{Callingham2024,
       author = {{Callingham}, J. R. and {Pope}, B. J. S. and {Kavanagh}, R. D. and others},
        title = "{Radio signatures of star–planet interactions, exoplanets and space weather}",
      journal = {\natas},
         year = 2024,
        month = nov,
       volume = {8},
       number = {11},
        pages = {1359-1372},
          doi = {10.1038/s41550-024-02405-6},
       adsurl = {https://ui.adsabs.harvard.edu/abs/2024NatAs...8.1359C},
}

@ARTICLE{Carson2013,
       author = {{Carson}, J. and {Thalmann}, C. and {Janson}, M. and others},
        title = "{Direct Imaging Discovery of a ``Super-Jupiter'' around the Late B-Type Star $\kappa$ And}",
      journal = {\apj},
         year = 2013,
        month = jan,
       volume = {763},
       number = {2},
        pages = {L32},
          doi = {10.1088/2041-8205/763/2/l32},
       adsurl = {https://ui.adsabs.harvard.edu/abs/2013ApJ...763L..32C},
}

@ARTICLE{Charbonneau2007,
       author = {{Charbonneau}, D. and {Brown}, T.~M. and {Burrows}, A. and others},
        title = "{When Extrasolar Planets Transit Their Parent Stars}",
      journal = {Protostars and Planets V},
         year = 2007,
        pages = {701-716},
       adsurl = {https://ui.adsabs.harvard.edu/abs/2007prpl.conf..701C},
}

@ARTICLE{Chen2017,
       author = {{Chen}, Jingjing and {Kipping}, David},
        title = "{PROBABILISTIC FORECASTING OF THE MASSES AND RADII OF OTHER WORLDS}",
      journal = {\apj},
         year = 2017,
        month = jan,
       volume = {834},
       number = {1},
        pages = {17},
          doi = {10.3847/1538-4357/834/1/17},
       adsurl = {https://ui.adsabs.harvard.edu/abs/2017ApJ...834...17C},
}

@ARTICLE{Chen2024,
       author = {{Chen}, Xuelei and {Gao}, Feng and {Wu}, Fengquan and others},
        title = "{Large-scale array for radio astronomy on the farside (LARAF)}",
      journal = {Philos. Trans. R. Soc. A},
         year = 2024,
        month = may,
       volume = {382},
        pages = {20230094},
          doi = {10.1098/rsta.2023.0094},
       adsurl = {https://ui.adsabs.harvard.edu/abs/2024RSPTA.38230094C},
}

@ARTICLE{Desch1984,
       author = {{Desch}, M. D. and {Kaiser}, M. L.},
        title = "{Predictions for Uranus from a radiometric Bode's law}",
      journal = {\nat},
         year = 1984,
        month = aug,
       volume = {310},
       number = {5980},
        pages = {755-757},
          doi = {10.1038/310755a0},
       adsurl = {https://ui.adsabs.harvard.edu/abs/1984Natur.310..755D},
}

@ARTICLE{Dempsey2021,
       author = {{Dempsey}, Adam M. and {Muñoz}, Diego J. and {Lithwick}, Yoram},
        title = "{Outward Migration of Super-Jupiters}",
      journal = {\apjl},
         year = 2021,
        month = sep,
       volume = {918},
       number = {2},
        pages = {L36},
          doi = {10.3847/2041-8213/ac22af},
       adsurl = {https://ui.adsabs.harvard.edu/abs/2021ApJ...918L..36D},
}

@ARTICLE{Driscoll2015,
       author = {{Driscoll}, P.E. and {Barnes}, R.},
        title = "{Tidal Heating of Earth-like Exoplanets around M Stars: Thermal, Magnetic, and Orbital Evolutions}",
      journal = {Astrobiology},
         year = 2015,
        month = sep,
       volume = {15},
       number = {9},
        pages = {739-760},
          doi = {10.1089/ast.2015.1325},
       adsurl = {https://ui.adsabs.harvard.edu/abs/2015AsBio..15..739D},
}

@ARTICLE{Elekes2023,
       author = {{Elekes}, F. and {Saur}, J.},
        title = "{Space environment and magnetospheric Poynting fluxes of the exoplanet $\tau$ Bo{\"o}tis b}",
      journal = {\aap},
         year = 2023,
        month = mar,
       volume = {671},
        pages = {A133},
          doi = {10.1051/0004-6361/202244947},
       adsurl = {https://ui.adsabs.harvard.edu/abs/2023A&A...671A.133E},
}

@ARTICLE{Farrell1999,
       author = {{Farrell}, W.~M. and {Desch}, M.~D. and {Zarka}, P.},
        title = "{On the possibility of coherent cyclotron emission from extrasolar planets}",
      journal = {\jgr},
         year = 1999,
       volume = {104},
        pages = {14025-14032},
          doi = {10.1029/1999JE900006},
       adsurl = {https://ui.adsabs.harvard.edu/abs/1999JGR...10414025F},
}

@ARTICLE{Farrell2003,
       author = {{Farrell}, W.~M. and {Desch}, M.~D. and {Lazio}, T.~J.~W.},
        title = "{The Potential of Radio Observations for Solar System and Extrasolar Planet Studies}",
      journal = {ASP Conf. Ser.},
         year = 2003,
       volume = {294},
        pages = {151},
       adsurl = {https://ui.adsabs.harvard.edu/abs/2003ASPC..294..151F},
}

@ARTICLE{Gasperin2020,
       author = {{de Gasperin}, F. and {Lazio}, T.~J.~W. and {Knapp}, M.},
        title = "{Radio observations of HD 80606 near planetary periastron -- II. LOFAR low band antenna observations at 30--78 MHz}",
      journal = {\aap},
         year = 2020,
       volume = {644},
        pages = {A157},
          doi = {10.1051/0004-6361/202038746},
       adsurl = {https://ui.adsabs.harvard.edu/abs/2020A&A...644A.157D},
}

@ARTICLE{Griemeier2005a,
       author = {{Grießmeier}, J.-M. and {Motschmann}, U. and {Mann}, G. and others},
        title = "{The influence of stellar wind conditions on the detectability of planetary radio emissions}",
      journal = {\aap},
         year = 2005,
        month = jul,
       volume = {437},
       number = {2},
        pages = {717-726},
          doi = {10.1051/0004-6361:20041976},
       adsurl = {https://ui.adsabs.harvard.edu/abs/2005A&A...437..717G},
}

@ARTICLE{Griemeier2007,
       author = {{Grie{\ss}meier}, J.-M. and {Zarka}, P. and {Spreeuw}, H.},
        title = "{Predicting low-frequency radio fluxes of known extrasolar planets}",
      journal = {\aap},
         year = 2007,
        month = nov,
       volume = {475},
       number = {1},
        pages = {359-368},
          doi = {10.1051/0004-6361:20077397},
       adsurl = {https://ui.adsabs.harvard.edu/abs/2007A&A...475..359G},
}

@ARTICLE{Griemeier2017,
       author = {{Grie{\ss}meier}, J.-M.},
        title = "{The radio emissions of exoplanets}",
      journal = {Planetary Radio Emissions VIII},
         year = 2017,
        pages = {285},
       adsurl = {https://ui.adsabs.harvard.edu/abs/2017pre8.conf..285G},
}

@ARTICLE{Hallinan2013,
       author = {{Hallinan}, G. and {Sirothia}, S. K. and {Antonova}, A. and others},
        title = "{Looking for a Pulse: A Search for Rotationally Modulated Radio Emission from the Hot Jupiter, $\tau$ Bo{\"o}tis b}",
      journal = {\apj},
         year = 2013,
        month = jan,
       volume = {762},
       number = {1},
        pages = {34},
          doi = {10.1088/0004-637x/762/1/34},
       adsurl = {https://ui.adsabs.harvard.edu/abs/2013ApJ...762...34H},
}

@ARTICLE{Hess2011,
       author = {{Hess}, S.~L.~G. and {Zarka}, P.},
        title = "{Modeling the radio signature of the orbital parameters, rotation, and magnetic field of exoplanets}",
      journal = {\aap},
         year = 2011,
       volume = {531},
        pages = {A29},
          doi = {10.1051/0004-6361/201116555},
       adsurl = {https://ui.adsabs.harvard.edu/abs/2011A&A...531A..29H},
}

@ARTICLE{Ip2004,
       author = {{Ip}, W.-H. and {Kopp}, A. and {Hu}, J.-H.},
        title = "{On the Star-Magnetosphere Interaction of Close-in Exoplanets}",
      journal = {\apjl},
         year = 2004,
       volume = {602},
        pages = {L53-L56},
          doi = {10.1086/382213},
       adsurl = {https://ui.adsabs.harvard.edu/abs/2004ApJ...602L..53I},
}

@ARTICLE{Lammer2007,
       author = {{Lammer}, Helmut and {Lichtenegger}, Herbert I.M. and {Kulikov}, Yuri N. and others},
        title = "{Coronal Mass Ejection (CME) Activity of Low Mass M Stars as An Important Factor for The Habitability of Terrestrial Exoplanets. II. CME-Induced Ion Pick Up of Earth-like Exoplanets in Close-In Habitable Zones}",
      journal = {Astrobiology},
         year = 2007,
        month = feb,
       volume = {7},
       number = {1},
        pages = {185-207},
          doi = {10.1089/ast.2006.0128},
       adsurl = {https://ui.adsabs.harvard.edu/abs/2007AsBio...7..185L},
}

@ARTICLE{Lazio2004,
       author = {{Lazio}, T.~J.~W. and {Farrell}, W.~M. and {Dietrick}, J.},
        title = "{The Radiometric Bode's Law and Extrasolar Planets}",
      journal = {\apj},
         year = 2004,
       volume = {612},
        pages = {511-518},
          doi = {10.1086/422552},
       adsurl = {https://ui.adsabs.harvard.edu/abs/2004ApJ...612..511L},
}

@ARTICLE{Lecavelier2009,
       author = {{Lecavelier Des Etangs}, A. and {Sirothia}, S.~K. and {Gopal-Krishna} and others},
        title = "{GMRT radio observations of the transiting extrasolar planet HD 189733 b at 244 and 614 MHz}",
      journal = {\aap},
         year = 2009,
       volume = {500},
        pages = {L51-L54},
          doi = {10.1051/0004-6361/200912347},
       adsurl = {https://ui.adsabs.harvard.edu/abs/2009A&A...500L..51L},
}

@ARTICLE{Lo2015,
       author = {{Lovis}, C. and {Fischer}, D.},
        title = "{Radial Velocity Techniques for Exoplanets}",
      journal = {Exoplanets (Univ. Arizona Press)},
         year = 2010,
        pages = {27-53},
}

@ARTICLE{Lynch2017,
       author = {{Lynch}, C.~R. and {Murphy}, T. and {Kaplan}, D.~L. and others},
        title = "{A search for circularly polarized emission from young exoplanets}",
      journal = {\mnras},
         year = 2017,
       volume = {467},
        pages = {3447-3453},
          doi = {10.1093/mnras/stx354},
       adsurl = {https://ui.adsabs.harvard.edu/abs/2017MNRAS.467.3447L},
}

@ARTICLE{Louarn1992,
       author = {{Louarn}, P.},
        title = "{Auroral planetary radio emissions, theoretical aspects}",
      journal = {Advances in Space Research},
         year = 1992,
        month = aug,
       volume = {12},
       number = {8},
        pages = {121-134},
          doi = {10.1016/0273-1177(92)90385-b},
       adsurl = {https://ui.adsabs.harvard.edu/abs/1992AdSpR..12..121L},
}

@ARTICLE{Louis2025,
       author = {{Louis}, C. K. and {Loh}, A. and {Zarka}, P. and others},
        title = "{Periodicities in radio emissions from Jupiter’s magnetosphere and consequences for radio emissions from star–exoplanet systems}",
      journal = {\aap},
         year = 2025,
        month = mar,
       volume = {707},
        pages = {A125},
          doi = {10.1051/0004-6361/202555183},
archivePrefix = {arXiv},
       eprint = {2503.18733},
 primaryClass = {astro-ph.EP},
       adsurl = {https://ui.adsabs.harvard.edu/abs/2025arXiv250318733L},
}

@ARTICLE{Majid2006,
       author = {{Majid}, W. and {Winterhalter}, D. and {Lazio}, J.},
        title = "{A Search for Radio Emission from Extrasolar Planets}",
      journal = {AAS Meeting Abstracts},
         year = 2006,
        pages = {90.05},
       adsurl = {https://ui.adsabs.harvard.edu/abs/2006AAS...209.9005M},
}

@ARTICLE{Mayor1995,
       author = {{Mayor}, Michel and {Queloz}, Didier},
        title = "{A Jupiter-mass companion to a solar-type star}",
      journal = {\nat},
         year = 1995,
        month = nov,
       volume = {378},
       number = {6555},
        pages = {355-359},
          doi = {10.1038/378355a0},
       adsurl = {https://ui.adsabs.harvard.edu/abs/1995Natur.378..355M},
}

@ARTICLE{Mayor2003,
       author = {{Mayor}, M.},
        title = "{Setting New Standards with HARPS}",
      journal = {The Messenger},
         year = 2003,
       volume = {114},
        pages = {20},
       adsurl = {https://ui.adsabs.harvard.edu/abs/2003Msngr.114...20M},
}

@ARTICLE{Melrose1982,
       author = {{Melrose}, D.~B. and {Dulk}, G.~A.},
        title = "{Electron-cyclotron masers as the source of certain solar and stellar radio bursts}",
      journal = {\apj},
         year = 1982,
       volume = {259},
        pages = {844-858},
          doi = {10.1086/160219},
       adsurl = {https://ui.adsabs.harvard.edu/abs/1982ApJ...259..844M},
}

@ARTICLE{MousaviSadr2025,
       author = {{Mousavi-Sadr}, Mahdiyar and {Tabatabaei}, Fatemeh S. and {Wolszczan}, Alexander and others},
        title = "{Exoplanetary radio emission predictions and detectability in the SKA era}",
      journal = {arXiv e-prints},
         year = 2025,
archivePrefix = {arXiv},
       eprint = {2510.19776},
 primaryClass = {astro-ph.EP},
       adsurl = {https://ui.adsabs.harvard.edu/abs/2025arXiv251019776M},
}

@ARTICLE{Murphy2015,
       author = {{Murphy}, T. and {Bell}, M.~E. and {Kaplan}, D.~L. and others},
        title = "{Limits on low frequency radio emission from southern exoplanets with the Murchison Widefield Array}",
      journal = {\mnras},
         year = 2015,
       volume = {446},
        pages = {2560-2565},
          doi = {10.1093/mnras/stu2253},
       adsurl = {https://ui.adsabs.harvard.edu/abs/2015MNRAS.446.2560M},
}

@ARTICLE{Li2023,
       author = {{Li}, Baoda and {Zhang}, Liyun and {Su}, Tianhao and {Han}, Xianming L. and {Misra}, Prabhakar and {Long}, Liu},
        title = "{Statistical and Radio Analysis of Exoplanets and Their Host Stars}",
      journal = {Universe},
         year = 2023,
       volume = {9},
       number = {11},
        pages = {475},
          doi = {10.3390/universe9110475},
}

@ARTICLE{Narang2024,
       author = {{Narang}, Mayank and {Manoj}, P. and {Chandra}, C.~H. Ishwara and others},
        title = "{A uGMRT search for radio emission from planets around evolved stars}",
      journal = {\mnras},
         year = 2024,
       volume = {529},
        pages = {1161-1168},
          doi = {10.1093/mnras/stae536},
       adsurl = {https://ui.adsabs.harvard.edu/abs/2024MNRAS.529.1161N},
}

@ARTICLE{Pineda2023,
       author = {{Pineda}, J. Sebastian and {Villadsen}, Jackie},
        title = "{Coherent radio bursts from known M-dwarf planet-host YZ Ceti}",
      journal = {\natas},
         year = 2023,
        month = apr,
       volume = {7},
       number = {5},
        pages = {569-578},
          doi = {10.1038/s41550-023-01914-0},
       adsurl = {https://ui.adsabs.harvard.edu/abs/2023NatAs...7..569P},
}

@ARTICLE{Ricker2015,
       author = {{Ricker}, George R. and {Winn}, Joshua N. and {Vanderspek}, Roland and others},
        title = "{Transiting Exoplanet Survey Satellite}",
      journal = {Journal of Astronomical Telescopes, Instruments, and Systems},
         year = 2015,
        month = oct,
       volume = {1},
       number = {1},
        pages = {014003},
          doi = {10.1117/1.jatis.1.1.014003},
       adsurl = {https://ui.adsabs.harvard.edu/abs/2015JATIS...1a4003R},
}

@ARTICLE{Ryabov2004,
       author = {{Ryabov}, V.~B. and {Zarka}, P. and {Ryabov}, B.~P.},
        title = "{Search of exoplanetary radio signals in the presence of strong interference: enhancing sensitivity by data accumulation}",
      journal = {\planss},
         year = 2004,
       volume = {52},
        pages = {1479-1491},
          doi = {10.1016/j.pss.2004.09.021},
       adsurl = {https://ui.adsabs.harvard.edu/abs/2004P&SS...52.1479R},
}

@ARTICLE{Seager2007,
       author = {{Seager}, S. and {Kuchner}, M. and {Hier‐Majumder}, C. A. and others},
        title = "{Mass‐Radius Relationships for Solid Exoplanets}",
      journal = {\apj},
         year = 2007,
        month = nov,
       volume = {669},
       number = {2},
        pages = {1279-1297},
          doi = {10.1086/521346},
       adsurl = {https://ui.adsabs.harvard.edu/abs/2007ApJ...669.1279S},
}

@ARTICLE{Shiratori2006,
       author = {{Shiratori}, Y.},
        title = "{Search for radio emissions from extrasolar planets}",
      journal = {36th COSPAR Scientific Assembly},
         year = 2006,
        pages = {2429},
       adsurl = {https://ui.adsabs.harvard.edu/abs/2006cosp...36.2429S},
}

@ARTICLE{Showman2010,
       author = {{Showman}, A.~P. and {Cho}, J.~Y-K. and {Menou}, K.},
        title = "{Atmospheric Circulation of Exoplanets}",
      journal = {Exoplanets},
         year = 2010,
        pages = {471-516},
       adsurl = {https://ui.adsabs.harvard.edu/abs/2010exop.book..471S},
}

@ARTICLE{Sirothia2014,
       author = {{Sirothia}, S.~K. and {Lecavelier des Etangs}, A. and {Gopal-Krishna} and others},
        title = "{Search for 150 MHz radio emission from extrasolar planets in the TIFR GMRT Sky Survey}",
      journal = {\aap},
         year = 2014,
        month = feb,
       volume = {562},
        pages = {A108},
          doi = {10.1051/0004-6361/201321571},
       adsurl = {https://ui.adsabs.harvard.edu/abs/2014A&A...562A.108S},
}

@ARTICLE{Smith2009,
       author = {{Smith}, A.~M.~S. and {Collier Cameron}, A. and {Greaves}, J. and others},
        title = "{Secondary radio eclipse of the transiting planet HD 189733 b: an upper limit at 307-347 MHz}",
      journal = {\mnras},
         year = 2009,
       volume = {395},
        pages = {335-341},
          doi = {10.1111/j.1365-2966.2009.14510.x},
       adsurl = {https://ui.adsabs.harvard.edu/abs/2009MNRAS.395..335S},
}

@ARTICLE{Stassun2018,
       author = {{Stassun}, Keivan G. and {Oelkers}, Ryan J. and {Pepper}, Joshua and others},
        title = "{The TESS Input Catalog and Candidate Target List}",
      journal = {\aj},
         year = 2018,
        month = sep,
       volume = {156},
       number = {3},
        pages = {102},
          doi = {10.3847/1538-3881/aad050},
       adsurl = {https://ui.adsabs.harvard.edu/abs/2018AJ....156..102S},
}

@ARTICLE{Turner2019,
       author = {{Turner}, J.~D. and {Grie{\ss}meier}, J.-M. and {Zarka}, P. and others},
        title = "{The search for radio emission from exoplanets using LOFAR beam-formed observations: Jupiter as an exoplanet}",
      journal = {\aap},
         year = 2019,
       volume = {624},
        pages = {A40},
          doi = {10.1051/0004-6361/201832848},
}

@ARTICLE{Turner2021,
       author = {{Turner}, J.~D. and {Zarka}, P. and {Grie{\ss}meier}, J.-M. and others},
        title = "{The search for radio emission from the exoplanetary systems 55 Cancri, $\upsilon$ Andromedae, and $\tau$ Bo{\"o}tis using LOFAR beam-formed observations}",
      journal = {\aap},
         year = 2021,
       volume = {645},
        pages = {A59},
          doi = {10.1051/0004-6361/201937201},
       adsurl = {https://ui.adsabs.harvard.edu/abs/2021A&A...645A..59T},
}

@ARTICLE{Turner2024,
       author = {{Turner}, J.~D. and {Zarka}, P. and {Grie{\ss}meier}, J.-M. and others},
        title = "{Follow-up LOFAR observations of the $\tau$ Bo{\"o}tis exoplanetary system}",
      journal = {\aap},
         year = 2024,
       volume = {688},
        pages = {A66},
          doi = {10.1051/0004-6361/202450095},
       adsurl = {https://ui.adsabs.harvard.edu/abs/2024A&A...688A..66T},
}

@ARTICLE{Vidotto2013,
       author = {{Vidotto}, A.~A. and {Jardine}, M. and {Morin}, J. and others},
        title = "{Effects of M dwarf magnetic fields on potentially habitable planets}",
      journal = {\aap},
         year = 2013,
       volume = {557},
        pages = {A67},
          doi = {10.1051/0004-6361/201321504},
       adsurl = {https://ui.adsabs.harvard.edu/abs/2013A&A...557A..67V},
}

@ARTICLE{Vidotto2020,
       author = {{Vidotto}, A.~A. and {Cleary}, A.},
        title = "{Stellar wind effects on the atmospheres of close-in giants: a possible reduction of escape and lack of strong atmospheric features}",
      journal = {\mnras},
         year = 2020,
       volume = {494},
        pages = {2417-2428},
          doi = {10.1093/mnras/staa911},
       adsurl = {https://ui.adsabs.harvard.edu/abs/2020MNRAS.494.2417V},
}

@ARTICLE{Wu1979,
       author = {{Wu}, C.~S. and {Lee}, L.~C.},
        title = "{A theory of the terrestrial kilometric radiation}",
      journal = {\apj},
         year = 1979,
       volume = {230},
        pages = {621-626},
          doi = {10.1086/157120},
       adsurl = {https://ui.adsabs.harvard.edu/abs/1979ApJ...230..621W},
}

@ARTICLE{Zarka1997,
       author = {{Zarka}, P. and {Queinnec}, J. and {Ryabov}, B.~P. and others},
        title = "{Ground-based high sensitivity radio astronomy at decameter wavelengths}",
      journal = {\planss},
         year = 1997,
       volume = {45},
        pages = {507},
}

@ARTICLE{Zarka1998a,
       author = {{Zarka}, Philippe},
        title = "{Auroral radio emissions at the outer planets: Observations and theories}",
      journal = {Journal of Geophysical Research: Planets},
         year = 1998,
        month = aug,
       volume = {103},
       number = {E9},
        pages = {20159-20194},
          doi = {10.1029/98je01323},
       adsurl = {https://ui.adsabs.harvard.edu/abs/1998JGR...10320159Z},
}

@ARTICLE{Zarka2001,
       author = {{Zarka}, Philippe and {Treumann}, Rudolf A. and {Ryabov}, Boris P. and others},
        title = "{Magnetically-Driven Planetary Radio Emissions and Application to Extrasolar Planets}",
      journal = {Astrophysics and Space Science},
         year = 2001,
       volume = {277},
        pages = {293-300},
          doi = {10.1007/978-94-010-0904-1_35},
       adsurl = {https://ui.adsabs.harvard.edu/abs/2001Ap&SS.277..293Z},
}

@ARTICLE{Zarka2004,
       author = {{Zarka}, P.},
        title = "{Radio and plasma waves at the outer planets}",
      journal = {Advances in Space Research},
         year = 2004,
       volume = {33},
        pages = {2045-2060},
          doi = {10.1016/j.asr.2003.07.055},
       adsurl = {https://ui.adsabs.harvard.edu/abs/2004AdSpR..33.2045Z},
}

@ARTICLE{Zarka2007,
       author = {{Zarka}, P.},
        title = "{Plasma interactions of exoplanets with their parent star and associated radio emissions}",
      journal = {\planss},
         year = 2007,
       volume = {55},
        pages = {598-617},
          doi = {10.1016/j.pss.2006.05.045},
       adsurl = {https://ui.adsabs.harvard.edu/abs/2007P&SS...55..598Z},
}

@ARTICLE{Zarka2014,
       author = {{Zarka}, Philippe and {Lazio}, Joseph and {Hallinan}, Gregg and others},
        title = "{Magnetospheric Radio Emissions from Exoplanets with the SKA}",
      journal = {PoS(AASKA14)},
         year = 2015,
        month = may,
        pages = {120},
          doi = {10.22323/1.215.0120},
       adsurl = {https://ui.adsabs.harvard.edu/abs/2015aska.confE.120Z},
}

@ARTICLE{Zarka2018,
       author = {{Zarka}, P. and {Marques}, M. S. and {Louis}, C. and others},
        title = "{Jupiter radio emission induced by Ganymede and consequences for the radio detection of exoplanets}",
      journal = {\aap},
         year = 2018,
        month = oct,
       volume = {618},
        pages = {A84},
          doi = {10.1051/0004-6361/201833586},
       adsurl = {https://ui.adsabs.harvard.edu/abs/2018A&A...618A..84Z},
}

@ARTICLE{Zarka2019,
       author = {{Zarka}, P. and {Li}, D. and {Grie{\ss}meier}, J.-M. and others},
        title = "{Detecting exoplanets with FAST?}",
      journal = {Research in Astronomy and Astrophysics},
         year = 2019,
       volume = {19},
        pages = {023},
          doi = {10.1088/1674-4527/19/2/23},
}

@ARTICLE{Vanderburg2020,
       author = {{Vanderburg}, Andrew and {Rappaport}, Saul A. and {Xu}, Siyi and others},
        title = "{A giant planet candidate transiting a white dwarf}",
      journal = {Nature},
         year = 2020,
       volume = {585},
        pages = {363-367},
          doi = {10.1038/s41586-020-2713-y},
}

@ARTICLE{Parsons2025,
       author = {{Parsons}, S.~G. and {Brown}, A.~J. and {Casewell}, S.~L. and {Littlefair}, S.~P. and others},
        title = "{Two almost planetary mass survivors of common envelope evolution}",
      journal = {\mnras},
         year = 2025,
       volume = {537},
        pages = {2112},
       eprint = {2501.14333},
       archivePrefix = {arXiv},
       adsurl = {https://ui.adsabs.harvard.edu/abs/2025MNRAS.537.2112P},
}

@ARTICLE{Braun2019,
       author = {{Braun}, R. and {Bonaldi}, A. and {Bourke}, T. and others},
        title = "{Anticipated Performance of the Square Kilometre Array -- Phase 1 (SKA1)}",
      journal = {arXiv e-prints},
         year = 2019,
archivePrefix = {arXiv},
       eprint = {1912.12699},
 primaryClass = {astro-ph.IM},
       adsurl = {https://ui.adsabs.harvard.edu/abs/2019arXiv191212699B},
}

@ARTICLE{vanHaarlem2013,
       author = {{van Haarlem}, M.~P. and {Wise}, M.~W. and {Gunst}, A.~W. and others},
        title = "{LOFAR: The LOw-Frequency ARray}",
      journal = {Astronomy \& Astrophysics},
         year = 2013,
        month = aug,
       volume = {556},
        pages = {A2},
          doi = {10.1051/0004-6361/201220873},
       adsurl = {https://ui.adsabs.harvard.edu/abs/2013A&A...556A...2V},
}

@ARTICLE{Gupta2017,
       author = {{Gupta}, Y. and {Ajithkumar}, B. and {Kale}, H.~S. and others},
        title = "{The upgraded GMRT: opening new windows on the radio Universe}",
      journal = {Current Science},
         year = 2017,
        month = aug,
       volume = {113},
       number = {4},
        pages = {707-714},
          doi = {10.18520/cs/v113/i04/707-714},
       adsurl = {https://ui.adsabs.harvard.edu/abs/2017CurrS.113..707G},
}

@ARTICLE{Christiansen2025,
       author = {{Christiansen}, J.~L. and others},
        title = "{The NASA Exoplanet Archive}",
      journal = {NASA Exoplanet Science Institute},
         year = 2025
}

@ARTICLE{Gu2026,
  author  = {Gu, Jun-Hua and Guo, Quan and Dong, Liang and others},
  title   = {{The Next-Generation 21CMA Telescope: Design, Commissioning, and Instrumental Effects in an SKA-LFAA-like System}},
  journal = {Research in Astronomy and Astrophysics},
  year    = {2026},
  volume  = {26},
  number  = {9},
  doi     = {10.1088/1674-4527/ae64c6}
}

@ARTICLE{Hadden2014,
       author = {{Hadden}, Sam and {Lithwick}, Yoram},
        title = "{Densities and Eccentricities of 139 Kepler Planets from Transit Time Variations}",
      journal = {\apj},
         year = 2014,
        month = may,
       volume = {787},
       number = {1},
        pages = {80},
          doi = {10.1088/0004-637X/787/1/80},
       adsurl = {https://ui.adsabs.harvard.edu/abs/2014ApJ...787...80H},
}

@ARTICLE{Canas2023,
       author = {{Ca{\~n}as}, Caleb I. and {Bender}, Chad F. and {Mahadevan}, Suvrath and others},
        title = "{Characterization of Low-mass Companions to Kepler Objects of Interest Observed with APOGEE-N}",
      journal = {ApJS},
         year = 2023,
        month = apr,
       volume = {265},
       number = {2},
        pages = {50},
          doi = {10.3847/1538-4365/acbcbe},
       adsurl = {https://ui.adsabs.harvard.edu/abs/2023ApJS..265...50C},
}

\end{document}